\documentclass[acmsmall,screen]{acmart}

\usepackage{microtype}
\usepackage{booktabs}
\usepackage{multirow}
\usepackage{balance}
\usepackage{enumitem}
\usepackage{listings}
\usepackage{xcolor}
\usepackage{amsmath}
\usepackage{wrapfig}
\usepackage{cleveref}
\usepackage{array}
\usepackage{subcaption}

\usepackage{pifont}


\setlist{itemsep=2pt, topsep=2pt, parsep=0pt}

\setlist[itemize]{noitemsep,topsep=2pt,leftmargin=*}
\setlist[enumerate]{noitemsep,topsep=2pt,leftmargin=*}

\lstdefinestyle{code}{
  basicstyle=\ttfamily\small,
  columns=fullflexible,
  breaklines=true,
  frame=single,
  rulecolor=\color{black!20},
  xleftmargin=2pt,
  xrightmargin=2pt
}
\renewcommand\footnotetextcopyrightpermission[1]{}

\title{Compiling Bioinformatics Recurrences}

\begin{document}

\author{Bala Vinaithirthan}
\email{balavinaithirthan@stanford.edu}
\affiliation{%
  \institution{Stanford University}
  \city{Stanford}
  \state{California}
  \country{USA}
}

\author{Shiv Sundram}
\email{shiv1@stanford.edu}
\affiliation{%
  \institution{Stanford University}
  \city{Stanford}
  \state{California}
  \country{USA}
}

\author{Sneha Goenka}
\email{goenka@princeton.edu}
\affiliation{%
  \institution{Princeton University}
  \city{Princeton}
  \state{New Jersey}
  \country{USA}
}
\author{Fredrik Kjolstad}
\email{kjolstad@stanford.edu}
\affiliation{%
  \institution{Stanford University}
  \city{Stanford}
  \state{California}
  \country{USA}
  }
\begin{abstract}
  Many bioinformatics algorithms, such as sequence alignment and structure prediction, can be expressed as recurrence equations over a dynamic programming matrix. Efficient implementations of these algorithms for large-scale biological data often require changing the order in which matrix cells are calculated and pruning ineffectual regions of the matrix from consideration altogether, but these techniques typically complicate implementation. We introduce FILTR, a domain-specific language (DSL) and compiler framework for bioinformatics recurrences. FILTR keeps the core recurrence rules separate from the pruning and scheduling strategies, where pruning acts as an approximation to limit where in the DP matrix cells are computed, and scheduling determines the iteration order for how cells are explored. FILTR compiles these high-level descriptions into optimized C++ code that matches the performance of hand-tuned implementations while enabling rapid exploration of new heuristics. FILTR is competitive with hand-optimized sequence-alignment libraries, ranging from $0.95\times$ to $30\times$ faster across biological benchmarks.
\end{abstract}

\maketitle

\section{Introduction}
\label{sec:introduction}

Living organisms are built from sequences of recurring sub-components. DNA consists of repeating nucleotides, genes contain recurring sequences, proteins are chains of repeating amino acids, and larger structures like cells and tissues are composed of recurring molecular building blocks~\cite{Liao2023RepetitiveDNA,Ridder2013PatternBioinfo}. Hence, modeling bioinformatics problems is often reduced to identifying and comparing recurring sequences and structural patterns~\cite{Rigoutsos1998Teiresias,Xue2019FrequentPatternsBioseq,Tarozzi2022RecurrentPatterns}. Examples include comparing two DNA sequences to see how similar they are, predicting how an RNA folds~\cite{zuker2003mfold, nussinov1978, mccaskill1990partition}, identifying shared patterns across genes~\cite{pevzner2001eulerian,zerbino2008velvet}, and inferring evolutionary relationships between species~\cite{felsenstein1981,durbin1998,gusfield1997}.

Recurrences decompose sequence modeling problems into overlapping sub-problems. A recurrence equation defines a sequence where each value is calculated in terms of its preceding terms. A classic example from bioinformatics is the Smith--Waterman algorithm~\cite{smith1981} that determines local alignment by identifying similar regions between a reference and a query sequence. Another example is the Needleman--Wunsch algorithm~\cite{needleman1970} that finds the similarity between sequences of comparable length such as complete genes or proteins. Both are ubiquitous in practice and form the crux of critical computational pipelines~\citep{mckenna2010gatk, li2009bwa, langmead2012bowtie2, koren2017canu, kolmogorov2019flye, hadfield2018nextstrain}. Recurrences are typically implemented with dynamic programming (DP). In both Smith--Waterman and Needleman--Wunsch, the core operation is a two-dimensional DP that generates an optimal \textit{alignment} by computing a matrix of possible edits, in which individual DP cells track the minimum cost of replacing, deleting, or inserting characters to morph, or \emph{align}, the query string into the reference string.


Real-world deployment of bioinformatics sequence algorithms relies on both semantics-preserving performance optimization and non-semantics-preserving algorithmic approximation to scale to millions or, in the case of the human genome, billions of base pairs~\cite{nhgri_basepair,t2t2022}. Moreover, different application pipelines require different accuracy thresholds---for example, finding a mutation requires the exact differences, while finding the regions onto which subsequences map requires only approximate matching. Since the design space of dynamic programs in bioinformatics is large and varied, bioinformaticians have invested substantial effort in creating optimized variants of alignment algorithms. Variants may be designed to algorithmically prune large parts of the computation~\cite{Koerkamp2024APairwise, Doblas2025QuickEd, ukkonen1985, harris2007lastz, zhang2000greedy, marcosola2021wfa, marcosola2023wfa2, ZhangEtAl2019LinearFold, ZhangEtAl2020LinearPartition, CourtneyEtAl2025Memerna, Eddy2011HMMER3}, to make more efficient use of the underlying hardware~\cite{Pham2023BWAGPU, LiuZhang2024HWAccelSurvey, koliogeorgi2021fpga, liu2009cudasw, logan, suzuki2018adaptive, AguadoPuig2023WFAGPU, SosicSikic2017Edlib, GaoEtAl2021abPOA}, or a combination of both. The global alignment algorithm, for example, admits several variants, such as banded alignment where only the band around the primary diagonal is computed as this is the most relevant part~\cite{ukkonen1985}. Even for this particular \emph{pruned} variant, many important optimization decisions remain. For instance, computing the matrix row by row exposes little parallelism but exhibits good cache locality, whereas computing along antidiagonals (the diagonals running from the bottom-left to the top-right) enables parallelization and vectorization at the cost of poorer cache behavior~\cite{yelick2020parallelmotifs,farrar2007}.

Implementations of bioinformatics sequence algorithms that are widely used in practice are hand-written and bundled in large libraries of different variants for different situations~\cite{harris2007lastz, li2018minimap2, goenka2020segalign, daily2016parasail, li2009bwa, sarvavid, seqan}. These implementations entangle algorithmic implementation, optimizations, and pruning approaches and must typically be rewritten for new pruning approaches or new platforms. Researchers have also developed programming systems for recurrence problems. The closest to our work is Recuma~\cite{recuma} and Bellman's Gap~\cite{bellmangap}. Recuma compiles recurrences over dense and sparse arrays but lacks support for algorithmic approximation through pruning optimizations as well as the diagonal iteration pattern that enables vectorized implementations. Bellman's gap is a logic programming system whose focus is on reducing exponential search to polynomial algorithms. Like Recuma it does not support diagonal iteration or dynamic pruning, although it has limited support for pruning by placing a static band around the diagonal.

In this paper, we show how to compile a simple but expressive recurrence language to efficient C++ code that implements the machine-level optimizations and approximate pruning optimizations that are used in modern bioinformatics. The optimizations are specified with two distinct optimization languages: a semantics-preserving scheduling language that controls both the order of iteration and the order of result storage, and a semantics-breaking pruning language that controls what parts of the iteration space to prune, presumably because it does not sufficiently contribute to the result. By separating the specification of recurrences from the specification of their optimization, a user can write a recurrence once in a simple syntax and then optimize it for different machines and data sets.
Our technical contributions are:
\begin{itemize}
    \item A recurrence-based intermediate representation that can express a wide range of bioinformatics recurrences along with iteration and pruning optimizations.
    \item An iteration order scheduling language that supports shearing and value-based iteration (search).
    \item A pruning language that applies semantics-breaking heuristic optimizations.
    \item A compiler that applies the iteration order and pruning languages to the recurrence IR and produces efficient C++ code.
\end{itemize}
We implemented our ideas in the FILTR compiler for bioinformatics recurrences, which generates native C++ code. Compiled recurrences achieve performance speedups of 0.95$\times$ to 30$\times$ against recurrence compilers and hand-optimized, often hand-vectorized libraries. We also show that FILTR can explore combinations of optimizations to find the best configuration for a given dataset. FILTR's composable design further enables optimizations unsupported by existing systems: the same pruning program, written once, can be applied across recurrences to produce novel pruned variants that would otherwise require separate hand-written implementations.

\section{Motivating Example}
\label{sec:motivating-example}
Sequence alignment is a core algorithm in bioinformatics that captures how genomic sequences diverge across organisms or across evolutionary time, through three fundamental mutational events: substitution, insertion, and deletion. Regardless of the optimizations applied in a particular use case, the core computation is a series of edit distances expressed as the recurrence:
\[
F_{i,j} =
\min \begin{cases}
  F_{i-1,j-1} + [q_i \neq r_j], & \text{match or mismatch (substitution)} \\
  F_{i-1,j} + 1, & \text{deletion} \\
  F_{i,j-1} + 1, & \text{insertion},
\end{cases}
\]
where $[q_i \neq r_j]$ denotes an Iverson bracket whose value is $1$ when the query sequence $q$ and reference sequence $r$ mismatch (differ) and $0$ when they match.

\begin{wrapfigure}{r}{0.54\textwidth}
    \vspace{-0.5em}
  \begin{center}
    \includegraphics[width=0.54\textwidth]{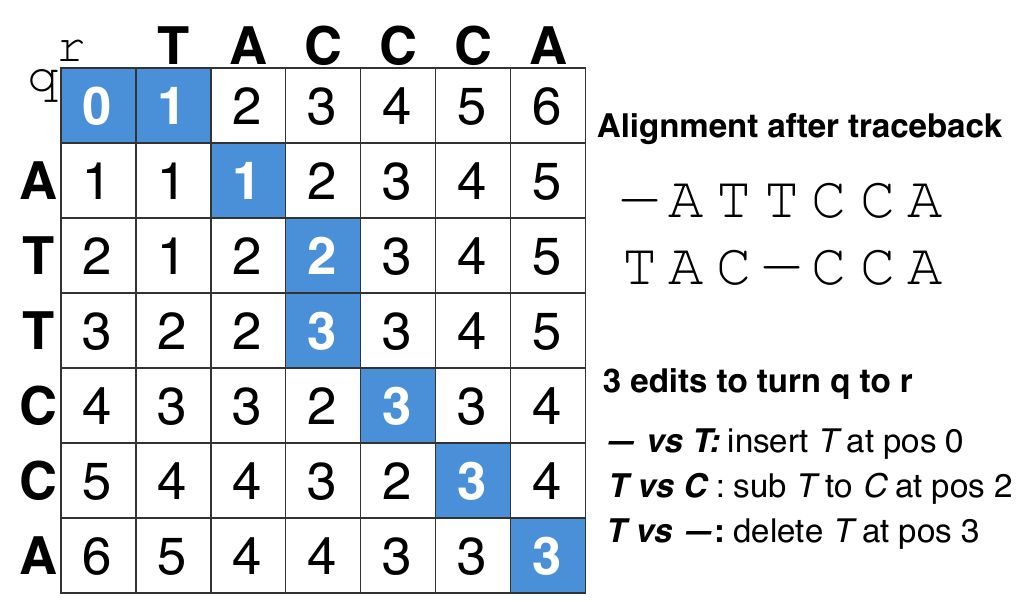}
  \end{center}
  \caption{Edit distance example.}
  \label{fig:edit-distance-formulation}
\end{wrapfigure}
\Cref{fig:edit-distance-formulation} shows the dynamic programming matrix $F$ that results from computing the edit distance. Each cell $(i,j)$ in the matrix represents the minimum cost of transforming the $1{\ldots}i$ prefix of the query string (labeled as q) on the vertical axis into the $1{\ldots}j$ prefix of the reference string (labeled as r) on the horizontal axis. As specified by the edit distance recurrence, this cost decomposes into three operations---a match/mismatch (diagonal move), an insertion (horizontal move), or a deletion (vertical move)---where each corresponds to a genetic variation: substituting, inserting, or deleting a single base pair (A, C, T, or G) in the sequence. 
Thus, each matrix entry or DP cell represents how the recurrence \emph{jointly arrives} at the prefixes $r_{1{\ldots}j}$ and $q_{1{\ldots}i}$ from smaller prefixes. The shaded cells and traceback path encode an optimal alignment---a sequence of three mutations (insertions, deletions, and substitutions) that transform the query string into the reference string, with the final edit distance equal to the number of mutations. An insertion corresponds to introducing a gap (denoted~``-'') in the query and a deletion corresponds to introducing a gap in the reference.

\begin{figure}
    \centering

    \begin{minipage}[b]{0.32\linewidth}
        \centering
        \includegraphics[width=\linewidth]{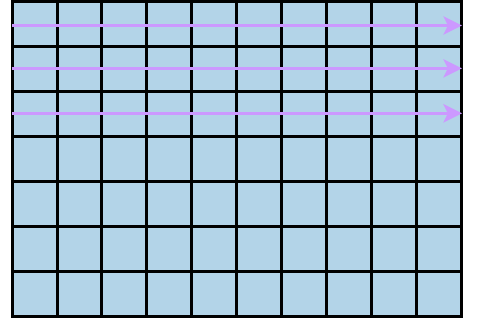}
        \subcaption{Standard alignment.}
        \label{fig:motivating-example-a}
    \end{minipage}
    \hfill
    \begin{minipage}[b]{0.32\linewidth}
        \centering
        \includegraphics[width=\linewidth]{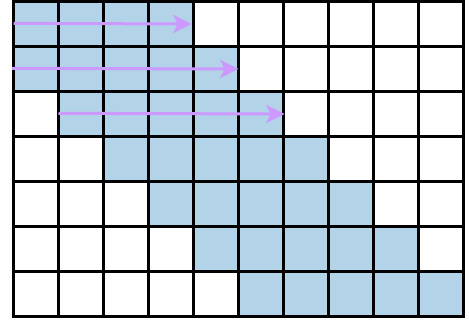}
        \subcaption{Banded alignment.}
        \label{fig:motivating-example-b}
    \end{minipage}
    \hfill
    \begin{minipage}[b]{0.32\linewidth}
        \centering
        \includegraphics[width=\linewidth]{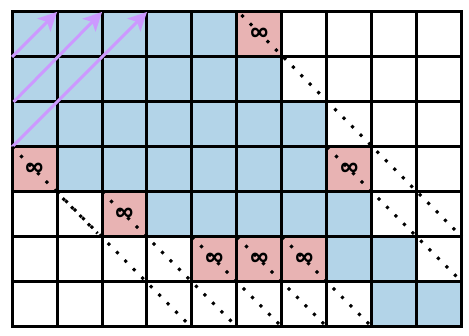}
        \subcaption{X-drop alignment.}
        \label{fig:motivating-example-c}
    \end{minipage}

    \caption{Three edit distance alignment variants that differ in where cells are computed and in what order. Colored cells indicate the active region. (a) Row-major standard alignment. (b) Banded alignment restricts computation to a static diagonal band. (c) X-drop alignment proceeds in antidiagonal order, marks cells as $\infty$ if they are unlikely, and prunes along the diagonal boundary established by $\infty$, producing data-dependent, dynamic pruning.}
    \label{fig:alignment-variant-comparison}
\end{figure}

This recurrence equation fully specifies the computation, yet few production aligners evaluate it directly on large sequences due to the significant cost. Real-world implementations optimize by controlling the order of computation and by pruning out work that is unlikely to contribute to the result. \Cref{fig:alignment-variant-comparison} shows three variants of the edit distance recurrence. These approaches share the same recurrence equation but have very different implementations due to different optimization strategies. The first variant (\Cref{fig:motivating-example-a}) is a standard vanilla implementation that executes the whole recurrence in row-major order.

The second variant (\Cref{fig:motivating-example-b}) is the banded alignment algorithm used in BWA-MEM~\cite{li2009bwa} and other production aligners~\cite{li2018minimap2, daily2016parasail, harris2007lastz, li2010bwasw}. Related biological sequences tend to align near the main diagonal of the dynamic programming matrix because large, sustained insertions or deletions are rare~\cite{mount2004bioinformatics}. Intuitively, aligning identical sequences would result in a path that lies exactly along the main diagonal of the matrix because there are no insertions and deletions. In this case, only diagonal cells would need to be computed. Assuming sequences are similar but not identical, only cells near the main diagonal must be explored. Banded alignment exploits this observation by restricting evaluation to a fixed-width diagonal band, as illustrated by the DP cells colored in \Cref{fig:motivating-example-b}.

The third variant (\Cref{fig:motivating-example-c}) is the X-drop alignment algorithm used in BLAST~\cite{altschul1997}, minimap2~\cite{li2018minimap2}, and others~\cite{logan, suzuki2018adaptive, harris2007lastz, seqan}. When alignments contain long stretches of insertions or deletions, such as the alignments found in evolutionary analysis between distantly related species, a fixed band may be too narrow to capture the optimal alignment or too wide to be fast. At the cost of more complicated control flow and data-dependent behavior, the X-drop algorithm improves on fixed banding by making the pruning data dependent. Rather than assuming a fixed diagonal bound, X-drop allows the alignment to dynamically determine which regions are promising based on the observed scores and which can be ignored. X-drop processes the alignment one antidiagonal at a time. As the recurrence advances, the algorithm tracks the best score seen on the previous antidiagonal. Any diagonal whose score falls more than a threshold $X$ below this score is marked inactive (shown as a red $\infty$ in \Cref{fig:motivating-example-c} and is not extended further along the diagonal (dotted line).

Methods such as above have been developed in bioinformatics to avoid evaluating the complete matrices, especially as the number of alignments~\cite{li2018minimap2} or length of the sequences increase significantly~\cite{harris2007lastz}. These make the computational time tractable while giving up minimal accuracy. As a result, most of the implementation effort goes into reducing the number of matrix cells evaluated and the order of evaluation, not writing the recurrence itself. This creates two problems. First, trying new pruning heuristics typically means reimplementing the whole algorithm, which discourages experimentation. Second, performance optimizations must be repeated for each pruning variant.

\section{The Recurrence Language}
\label{sec:recurrence-language}

Bioinformaticians reason about algorithms in terms of their recurrence equations. These equations are the shared language of the field, appearing in papers and lectures as the canonical description of how an algorithm models biological reality~\cite{needleman1970, durbin1998, smith1981, gusfield1997}. 

FILTR's recurrence language serves as the primary user-facing input to FILTR, capturing the mathematical structure of a bioinformatics model exactly as described in the literature, without implementation detail about how or where it is evaluated. FILTR's recurrence language is based on the Recuma language~\cite{recuma}, and follows the grammar described in~\Cref{fig:recurrence-language-grammar}.
The language is made up of one or more recurrence equations that each defines the value of a cell in a multidimensional array as a function of previously computed values, input data, and scalar constants. The function can be composed of arithmetic operators, aggregation operators such as $\max$, reductions over index ranges, and side-effect-free user-defined functions.

\begin{figure}[b]
\begin{minipage}[b]{0.41\linewidth}
 \begingroup
  \setlength{\arraycolsep}{2pt}
  \scriptsize
  \begin{minipage}[t]{\linewidth}
  \[
  \begin{array}{r@{\;}l@{\;}l}
    \textsf{Prog} & ::= 
      \textsf{Rec}^{+} \\[0.2em]
    \textsf{Rec} & ::= 
      \textsf{TA} = \textsf{Expr} \\[0.2em]
    \textsf{TA} & ::= 
      t[\textsf{Idx}(,\textsf{Idx})^{*}] \\[0.2em]
    \textsf{Idx} & ::= 
      v + n \quad (n \in \mathbb{Z}) \\[0.2em]
    \textsf{Expr} & ::= 
      \textsf{TA}
      \mid \textit{const} \\[0.2em]
      & \mid
      \textsf{Expr}\;\textsf{Op}\;\textsf{Expr} \\[0.2em]
      & \mid
      \textsf{ReduceOp}_{v,\;\textsf{Cond}^{*}}\;\textsf{Expr} \\[0.2em]
      & \mid
      f(\textsf{Expr}(,\textsf{Expr})^{*}) \\
    \textsf{Op} & ::= 
      + \mid - \mid * \mid / 
      \mid \max \mid \min \\[0.2em]
    \textsf{ReduceOp} & ::= 
      \Sigma \mid \Pi
      \mid \max \mid \min
      \mid \arg\max \mid \arg\min \\[0.2em]
    \textsf{Cond} & ::=
      \textsf{Expr}\;\textsf{CmpOp}\;\textsf{Expr} \\[0.2em]
    \textsf{CmpOp} & ::=
      {<} \mid {\le} \mid {>} \mid {\ge} \mid {=} \mid {\ne}
  \end{array}
  \]
  \end{minipage}
  \endgroup
  \caption{Grammar of recurrence language.}
  \label{fig:recurrence-language-grammar}
\end{minipage}
\hfill
\begin{minipage}[b]{0.55\linewidth}
\centering
\scriptsize
\renewcommand{\arraystretch}{1.3}
\setlength{\tabcolsep}{3pt}
\begin{tabular}{|m{1.1cm}|l|}
\hline
\textbf{Topic} & \textbf{Example Recurrence} \\
\hline
Sequence Alignment &
$\begin{aligned}
&M_{i,j} = \max(M_{i-1,j-1},\, I_{i-1,j-1},\, D_{i-1,j-1}) + \mathit{in1}_{i,j} \\
&I_{i,j} = \max(M_{i-1,j} - c_1,\; I_{i-1,j} - c_2) \\
&D_{i,j} = \max(M_{i,j-1} - c_1,\; D_{i,j-1} - c_2)
\end{aligned}$ \\
\hline
Sequence Chaining &
$C_j = \mathit{in1}_j + \max_{i<j} C_i$ \\
\hline
RNA \mbox{Folding} &
$\begin{aligned}
F_{i,j} = \max&(
F_{i+1,j},\,
F_{i,j-1},\,
F_{i+1,j-1}+\mathit{in1}_{i,j},\, \\
\max&_{i \le k < j}(F_{i,k}+F_{k+1,j})
)
\end{aligned}$ \\
\hline
Sequence Prediction &
$\begin{aligned}
&M_{i,j} = \mathit{in1}_{i,j} \cdot \max(M_{i-1,j-1} \cdot c_1,\, I_{i-1,j-1} \cdot c_2,\, D_{i-1,j-1} \cdot c_3) \\
&I_{i,j} = \mathit{in2}_{i} \cdot \max(M_{i-1,j} \cdot c_4,\, I_{i-1,j} \cdot c_5,\, D_{i-1,j} \cdot c_6) \\
&D_{i,j} = \mathit{in3}_{j} \cdot \max(M_{i,j-1} \cdot c_7,\, I_{i,j-1} \cdot c_8,\, D_{i,j-1} \cdot c_9)
\end{aligned}$ \\
\hline
\end{tabular}
\caption{Example recurrences for different biology problems.
}
\label{tab:dp_recurrences}
\end{minipage}
\end{figure}

\Cref{tab:dp_recurrences} shows representative recurrences from four major classes of bioinformatics problems. Though they model different biological processes, they share several structural properties. First, dependencies follow regular patterns: cells reference neighbors along rows, columns, diagonals, and antidiagonals. Second, many recurrences are mutually defined: affine-gap alignment (shown as Sequence Alignment example recurrence) and sequence prediction each couple three arrays ($M$, $I$, $D$) that reference one another at each step. Third, some recurrences involve reductions over variable-length ranges---RNA folding maximizes over $k$, and sequence chaining reduces over all predecessors. In all cases, a global optimization problem decomposes into overlapping subproblems evaluated over a rectangular or triangular index space.

\section{System Overview}
\label{sec:system-overview}

\begin{wrapfigure}{r}{0.55\textwidth}
  \begin{center}
    \vspace{-2em}
    \includegraphics[width=1\linewidth]{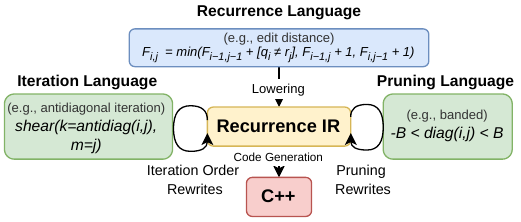}
    \end{center}
    \vspace{-0.3em}
    \caption{The recurrence language is lowered to a Recurrence IR, which is then optimized by user-specified iteration and pruning rewrites before C++ code generation. We include an antidiagonal-iterated banded edit distance example.}
    \label{fig:system-overview-simple}
\end{wrapfigure}

The FILTR system compiles simple recurrence equations to efficient C++ code. \Cref{fig:system-overview-simple} gives an overview of the compilation process. Recurrence equations expressed in the language of \Cref{sec:recurrence-language} are lowered to a recurrence intermediate representation (Recurrence IR), which is a more sophisticated recurrence language that can express the optimized iteration order and iteration pruning optimizations needed for state-of-the-art bioinformatics kernels. Two optimization languages (shown in green) are used to specify rewrites on the recurrence IR. While the Recurrence Language (shown in blue) specifies \emph{what} is computed, the Iteration Ordering Language specifies \emph{how} to order the computation, and the Pruning Language specifies \emph{where} computation should be skipped.

This separation of concern lets a user write down a bioinformatics recurrence using simple intuitive Cartesian coordinates and bounds, and then vary iteration order optimizations and pruning strategies depending on the characteristics of the target machine and sequence data. To illustrate, the base edit distance recurrence, using Cartesian coordinates, is $F_{i,j} = \min(F_{i-1,j-1} + [q_i \neq r_j],\ F_{i-1,j}+1,\ F_{i,j-1}+1)$ where $0\leq i \leq M$ and  $0\leq j \leq N$. It clearly indicates that $F_{i,j}$ is determined by three neighboring cells corresponding to insertion, deletion, and replacement/matching. The equivalent, antidiagonal form of this recurrence is $F_{k,m} = \min(F_{k-2,m-1} + [q_{k-m} \neq r_m],\ F_{k-1,m}+1,\ F_{k-1,m-1}+1)$ where $0 \leq k=i+j \leq M + N$ and $\max(0, k-M) \leq m=j \leq \min(N,k)$. The antidiagonal form (when expressed over loops over $k$ and $m$) is vectorizable and thus more performant, but more complex for users to express and understand. In FILTR, the user can provide the simpler version of edit distance in the Recurrence Language, and transform it into an antidiagonal form in the Recurrence IR (\Cref{sec:recurrence-IR}) with the command $shear(k=antidiag(i,j), m=j)$ from the Iteration Ordering Language (\Cref{fig:system-overview-simple}).

The \emph{Iteration Ordering Language} (\Cref{sec:iteration-ordering}) lets the user control how a recurrence is executed by specifying the order in which its matrix cells are traversed. It is a semantics-preserving language, which means it guarantees the same final values of the recurrence and the traceback path. As an optimization, the chosen traversal order also automatically determines the storage layout: cells are stored so that consecutively visited cells are contiguous in memory. Storage optimizations arise naturally from traversal strategy and in practice, bioinformatics workloads do not require storage and iteration to be specified independently as is common in, e.g., tensor algebra~\cite{kjolstad2017taco}.

The \emph{Pruning Language} (\Cref{sec:pruning-rewrites}) is a semantics-breaking approximation language that lets the user remove points in the recurrence's iteration domain from the computation. Users express constraints that define a region of the iteration domain and only result cells within this region are computed. These constraints may be simple inequalities over the iteration domain coordinates (static pruning) or they may take the form of recurrences over the iteration domain whose values evolve during execution (dynamic pruning).

\section{Recurrence IR}
\label{sec:recurrence-IR}

The Recurrence IR (RIR) is a language expressive enough to fully describe an optimized bioinformatics algorithm, including its recurrence logic, its traversal order, and pruning of the iteration domain. The Recurrence Language from \Cref{sec:recurrence-language} was deliberately designed to describe only the mathematical model, and to say nothing about how or where the recurrences are evaluated. This simplicity allows the Recurrence Language to express a large number of algorithms from the bioinformatics literature in a simple and intuitive way, but prohibits it from representing the banding, dynamic pruning, and non-standard iteration orders that optimized kernels require. 

\begin{wrapfigure}{r}{0.45\textwidth}
  \centering
  \begingroup
  \setlength{\arraycolsep}{2pt}
  \scriptsize
  \begin{minipage}[t]{\linewidth}
  \[
  \begin{array}{r@{\;}l@{\;}l}
    \textsf{RIR} & ::=
      \textsf{TA}\;:\;\textsf{Region}^{+} \\[0.2em]
    \textsf{Region} & ::=
      \textsf{Stmt}^{+}\;\textsf{domain}\;:\;\textsf{Cond}^{+} \\[0.2em]
    \textsf{Stmt} & ::=
      \textsf{Rec} \\[0.2em]
      & \mid
      \textsf{TA} = \textsf{Expr}\;\textsf{if}\;\textsf{Cond}\;\textsf{else}\;\textsf{Expr} \\[0.2em]
    \textsf{Expr} & ::=
      \ldots \quad \textit{(as in \Cref{fig:recurrence-language-grammar})} \\[0.2em]
      & \mid
      \textsf{ReduceOp}\,(\,\textsf{Expr } \textsf{if } \textsf{Cond}^{+}\,)
  \end{array}
  \]
  \end{minipage}
  \endgroup
  \caption{Grammar of Recurrence IR, extending the Recurrence Language (\Cref{fig:recurrence-language-grammar}). Each region carries its own domain conditions, enabling domain splitting. Domain bounds may be expressions over outer index variables and recursive.}
  \label{fig:rir-grammar}
\end{wrapfigure}

\begin{figure}[b]
    \centering
    \includegraphics[width=1\linewidth]{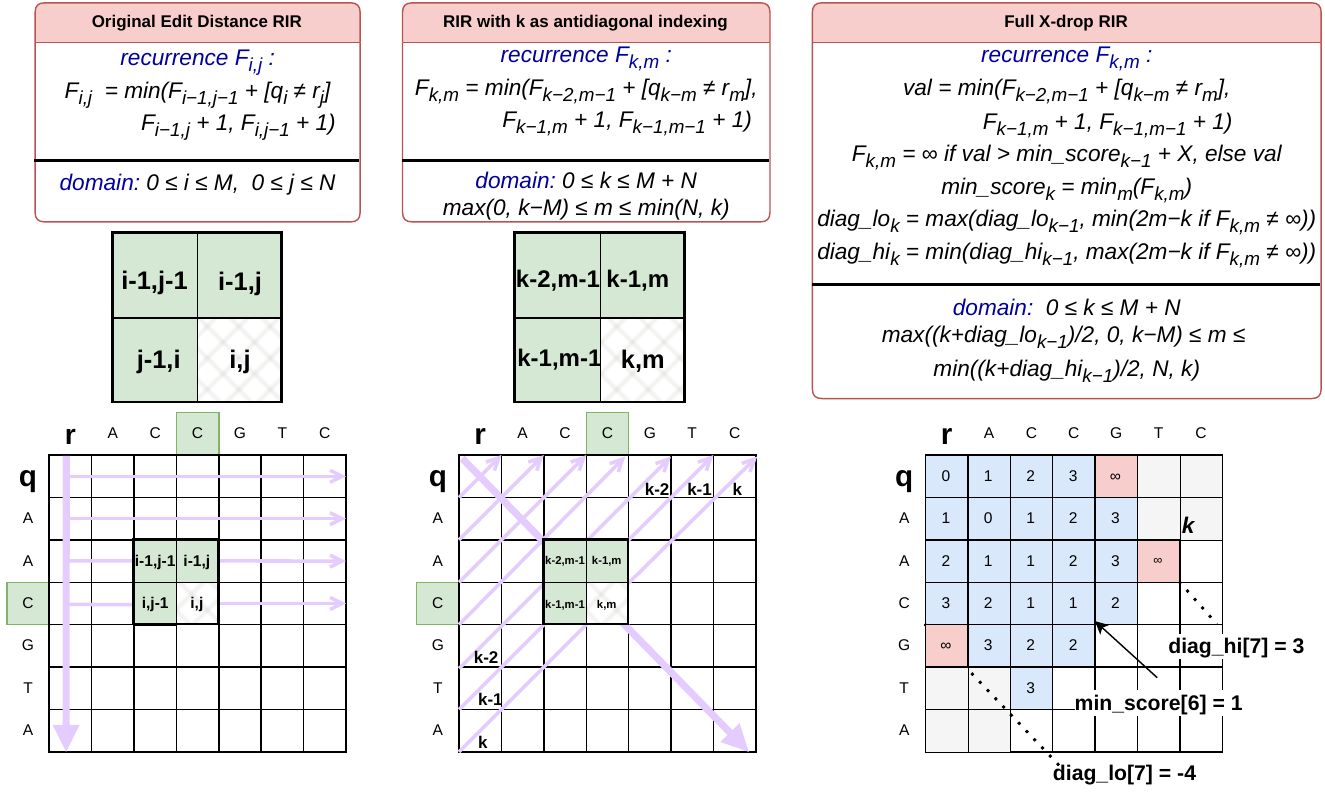}
\caption{Progressive refinement of the edit distance Recurrence IR. (Left) The original Cartesian formulation indexes by $(i, j)$ with row-major traversal. (Center) After changing to antidiagonal coordinates $(k, m)$, the same values are computed along antidiagonals. Dependencies are shown in green. (Right) The full X-drop formulation adds conditional and domain recurrences; the active region narrows as cells exceeding the score threshold are set to $\infty$ and the bounds $\text{diag\_lo}$, $\text{diag\_hi}$ contract accordingly.}    \label{fig:six-panel-RIR}
\end{figure}

The RIR extends the Recurrence Language with the grammar in \Cref{fig:rir-grammar} that introduces features to allow for exploration of iteration and pruning strategies---explicit domains and conditional definitions. The RIR allows domain bounds on iteration variables to be expressions over outer index variables and to be defined by their own recurrences. These allow the iteration space to vary with the outer loop index and to contract at runtime based on previously computed values. The RIR also supports conditionally defined recurrence equations and set-builder reductions. A recurrence equation may produce different values depending on a condition, and a reduction may aggregate over indices satisfying a condition. \Cref{fig:six-panel-RIR} illustrates these additional features through three progressively more complex formulations of edit distance.

An algorithm in the RIR consists of one or more recurrence equations that jointly compute one or more tensors. Each recurrence equation is paired with a domain that specifies the values over which its index variables range. When a recurrence is first lowered from the Recurrence Language, the compiler infers a simple domain from the index variables; the scheduling and pruning languages then embellish both the equations and the domain.

\subsection{Recurrence Domains}
\label{sec:recurrence-domains}
Every recurrence in the RIR is paired with a domain that specifies the bounds over which its index variables range. Domains in the RIR may be static (e.g., $-3 \!<\! i \!<\! 3$), or dynamic (e.g., $\mathrm{col\_lo_i} \!<\!j \!<\! \mathrm{col\_hi_i}$), where $\mathrm{col\_lo_i}$ and $\mathrm{col\_hi_i}$ are dynamically calculated (via separate recurrences) on each new row $i$.

In the simplest case---a rectangular domain---the bounds are constants. The original edit distance (\Cref{fig:six-panel-RIR}, left) is defined over $0 \le i \le M,\; 0 \le j \le N$. When a recurrence is first lowered from the Recurrence Language, the compiler infers constant bounds on each index variable, producing a simple domain. This works well for straightforward Cartesian formulations.

Many optimized algorithms require more complex iteration spaces, beyond the simple bounds that can be expressed with constants. The RIR thus allows bounds on inner index variables to be symbolic expressions over outer index variables. For instance, antidiagonal-major traversal, where cells are grouped by the sum $k = i + j$, produces a domain in which the number of cells per antidiagonal first grows (until hitting the middle antidiagonal), and then shrinks. The correct bounds, shown in \Cref{fig:six-panel-RIR} (center),  are $\max(0, k{-}M) \le m \le \min(N, k)$, where $m$ is the position along antidiagonal $k$. These derived bounds are the result of a Cartesian to diagonal coordinate transform that our compiler handles automatically (\Cref{sec:iteration-ordering}).

Some algorithms require the domain itself to change during execution. X-drop and other dynamic pruning strategies from literature contract the active region based on the actual scores that have been so far calculated. To support this, the RIR allows recurrences to determine the domain, not just the output data. More specifically, \emph{data recurrences} (the recurrences seen up to this point) like edit distance are merely evaluated over cells within a given domain. A \emph{domain recurrence}, on the other hand, is a recurrence whose computed values define the \emph{domain}---the bounds over which data recurrences are evaluated. This makes the iteration space itself programmable---what cells the algorithm computes next can depend on what it has computed so far. The X-drop formulation (\Cref{fig:six-panel-RIR}, right) illustrates this: $\mathrm{diag\_lo}_k$ and $\mathrm{diag\_hi}_k$ are domain recurrences that narrow the range of $m$ at each antidiagonal based on which cells were dynamically pruned on previous antidiagonals. The lower bound on $m$ incorporates $\mathrm{diag\_lo}_{k-1}$, and the upper bound incorporates $\mathrm{diag\_hi}_{k-1}$, so the active region shrinks as computation proceeds.

Domain recurrences and data recurrences may be mutually recursive, meaning their calculations are interdependent. In X-drop, the data recurrence for $F_{k,m}$ determines which cells are given tangible scores or set to $\infty$; the domain recurrences for $\mathrm{diag\_lo}$ and $\mathrm{diag\_hi}$ analyze the spatial distribution of these $\infty$ values on the current antidiagonal and then update the domain bounds, which define which cells of $F$ will be evaluated on the next antidiagonal. This mutual recursion---in which data values influence future domain bounds, and domain bounds determine which cells are computed---is exactly what enables data-dependent pruning at runtime. The Pruning Language (\Cref{sec:pruning-rewrites}) provides the user-facing constructs for specifying such strategies to be applied to the RIR.

\subsection{Recurrence Equations}
In the simplest case, a recurrence equation in the RIR directly computes an array element from previously computed values, exactly as in the Recurrence Language. The original edit distance in \Cref{fig:six-panel-RIR} (left) is an example. The RIR extends the Recurrence Language with conditional definitions and set-builder reductions.

A recurrence equation may be \emph{conditionally defined}: a single array element is given by one of several expressions, each guarded by a Boolean condition built from comparisons over index variables, constants, input data, or previously computed values. Simple index-only conditions (such as $i = 0$) are useful for representing base cases. More powerfully, conditions that depend on computed values enable pruning within the recurrence itself. 

The X-drop formulation in \Cref{fig:six-panel-RIR} (right) demonstrates this: $F_{k,m} = \infty$ if $\mathrm{val} > \mathrm{min\_score}_{k-1} + X$, else $\mathrm{val}$. If the candidate score exceeds the best score seen on the previous antidiagonal by more than the X-drop threshold, the cell is pruned by setting it to $\infty$. In \Cref{fig:six-panel-RIR}, the cells marked $\infty$ at antidiagonal $k{=}7$ are the result of this pruning. Otherwise, the cell keeps its computed value.

The RIR also adds a \emph{set-builder reduction}, which reduces an expression over all
indices satisfying a condition. For example, $\mathrm{diag\_lo}_k =
\max\!\bigl(
\mathrm{diag\_lo}_{k-1},\;
\min (\, 2m-k \;\text{if}\; F_{k,m} \ne \infty \,)
\bigr)$.
This finds the smallest diagonal index among non-pruned cells on antidiagonal~$k$. The analogous formula for $\mathrm{diag\_hi}_k$ uses $\max$ in place of $\min$ to find the largest active diagonal index. These values tighten---via the outer
$\max$ and $\min$ with the previous antidiagonal's bounds---to narrow the active region of the matrix. Together with the data recurrence tracking the $\mathrm{min\_score}$, these set-builder reductions complete the data-domain loop that allows for dynamic pruning.

\section{Iteration Ordering Language}
\label{sec:iteration-ordering}
A recurrence's iteration space is the geometric shape of its domain. Its iteration order is the order in which points in that space are visited. FILTR provides three index-space transformations---\textit{loop order}ing, \textit{shear}ing, and \textit{search}ing---whose grammar is given in \Cref{fig:iteration-language-grammar}. These transformations rewrite the recurrence's iteration order and, as an automatic optimization, its storage layout and iteration space. Choosing a traversal order and a storage layout are tightly intertwined decisions because the order in which cells are visited informs which values should be kept in memory and cache. Performance of a recurrence program depends on the order in which cells are evaluated: different traversal orders can expose parallelism, improve memory locality, or eliminate impossible states entirely.

\begin{wrapfigure}{r}{0.45\textwidth}
  \vspace{-1em}
  \centering

  \begingroup
  \setlength{\arraycolsep}{2pt}
  \scriptsize

  \begin{minipage}[t]{\linewidth}
  \[
  \begin{array}{r@{\;}l@{\;}l}
    \textsf{Prog} & ::= 
      \textsf{Trans}^{*} \\[0.2em]

    \textsf{Trans} & ::= 
      \textsf{LoopOrd}
      \mid \textsf{Shr}
      \mid \textsf{Search} \\[0.2em]

    \textsf{LoopOrd} & ::= 
      \textit{loopOrd}(\textsf{Vars}) \\[0.2em]

    \textsf{Shr} & ::= 
      \textit{shear}(\textsf{CoordDef}
      (,\textsf{CoordDef})^{*}) \\[0.2em]

    \textsf{CoordDef} & ::= 
      v=\textsf{CoordExpr} \\[0.2em]

    \textsf{Search} & ::= 
      \textit{search}(\textit{score},
      \textsf{CoordDef}
      (,\textsf{CoordDef})^{*}) \\[0.2em]

    \textsf{CoordExpr} & ::= 
      \textit{antiDiag}(\textsf{Vars})
      \mid \textit{diag}(\textsf{Vars})
      \mid v \\[0.2em]

    \textsf{Vars} & ::= 
      v(,v)^{*}
  \end{array}
  \]
  \end{minipage}

  \endgroup

  \caption{Grammar of Iteration Ordering Language. loopOrd, antidiag, and diag are abbreviations.}
  \label{fig:iteration-language-grammar}
  \vspace{-1em}
\end{wrapfigure}

Ordering commands are named rewrites on the RIR and leave the recurrence's traceback and final value unchanged, i.e.\, are correctness preserving. Each transformation takes an RIR as input and produces a rewritten RIR in a new coordinate system. \textit{loopOrd}, the simplest case, swaps the nesting of index variables. \textit{Shearing} introduces new coordinates as linear combinations of the originals, reshaping the iteration space so that independent computations become contiguous. This transformation both changes traversal order and reorders data layout to be antidiagonal major, as opposed to row major or column major. \textit{Search} converts the indices of the original matrix so that position becomes the stored quantity, and score or cost (i.e., edit distance) becomes an index, intuitively converting a DP matrix traversal into a search that visits only possible states. This transformation changes traversal and reorders data layout to be ``score-major''. We describe the user-facing specification of each transformation first, then the compiler rewrites that implement them.

\subsection{Loop Order}
\label{sec:loop-order}

A recurrence permits multiple valid evaluation orders. In edit distance, row-major traversal, iterating over $i$ in the outer loop and $j$ in the inner loop, visits cells row by row, producing good cache locality for row-major storage. This and other valid orderings compute the same values; however, the order in which the DP matrix is filled determines memory access patterns, parallelism, and compatibility with pruning strategies.

\paragraph{Language Feature}
The simplest iteration transformation reorders loop nesting. A recurrence over index variables $(i,j)$ admits multiple valid evaluation orders. The user specifies the desired nesting with the \textit{loopOrd} construct: \textit{loopOrd(i, j)} produces row-major traversal; \textit{loopOrd(j, i)} produces column-major.

\paragraph{RIR Rewrite}
Given a requested loop order, the compiler rewrites the domain by solving the bound constraints for each variable in nesting order: the outermost variable's bounds are solved as constants or expressions independent of inner variables, and each subsequent variable's bounds are solved in terms of the enclosing variables. For simple domains, this amounts to swapping the order of the bound expressions. For more complex domains produced by subsequent shearing transformations (\Cref{sec:shearing}), the compiler re-solves the symbolic inequalities in the new variable order. After rewriting the domain, the compiler verifies that the loop order respects the recurrence's data dependencies. If a user specifies a loop ordering that violates a dependency, the compiler rejects the program with an error.

\subsection{Shearing}
\label{sec:shearing}

\begin{wrapfigure}{r}{0.55\textwidth}
  \begin{center}
    \includegraphics[width=1\linewidth]{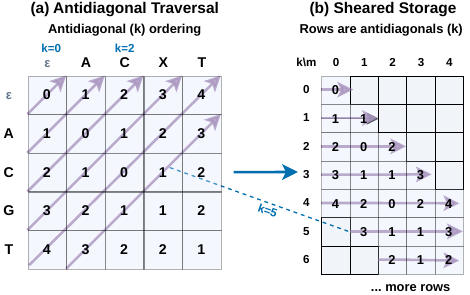}
    \end{center}
    \caption{(a, left) Cells on the same antidiagonal (constant i+j) are independent but lie in different rows. (b, right) After shearing, each antidiagonal maps to a contiguous row in the storage layout. Shearing couples traversal and storage (antidiagonal becomes sequential in memory).}
    \label{fig:antidiagonal-shearing}
\end{wrapfigure}

In the edit distance recurrence, neither rows nor columns offer straightforward parallelism. Each cell $(i, j)$ in the edit distance recurrence depends on $(i-1, j)$, $(i, j-1)$, and $(i-1, j-1)$, neighbors in both the row and column directions. Iterating along a row, every cell depends on the cell immediately to its left; iterating down a column, every cell depends on the cell immediately above it. In either case, cells within the same loop iteration are linked by dependencies, preventing straightforward parallel evaluation.

However, all three dependencies share a property: they have strictly smaller values of $i + j$. This means that every cell on a given antidiagonal---the set of cells where $i + j$ equals some constant $k$---depends only on cells from earlier antidiagonals. All cells on the same antidiagonal are mutually independent and can be computed in parallel (\Cref{fig:antidiagonal-shearing}).

Antidiagonal traversal solves the parallelism problem but introduces a memory locality problem. If the DP matrix is stored in row-major order, consecutive cells on an antidiagonal lie in different rows and typically in different cache lines. Each step along the antidiagonal jumps to a new row, producing strided memory accesses. Shearing addresses both problems by coupling storage and traversal: it reshapes the iteration space so that antidiagonals are traversed in sequence and, as a direct result, reshapes storage so that each antidiagonal occupies a contiguous block in memory. After shearing, the cells of each antidiagonal are contiguous in memory, so parallel traversal benefits from good locality. \Cref{fig:antidiagonal-shearing} illustrates this: panel (a) shows the strided access pattern of antidiagonal traversal over the original matrix, while panel (b) shows the sheared storage layout where each antidiagonal $k$ maps to a contiguous row.

\paragraph{Language Feature}

The user requests a shearing transformation using the \textit{shear} command, where the new coordinates are defined as linear combinations of the original coordinates. For antidiagonal storage in edit distance, the user writes $\textit{shear} (k = \textit{antidiag}(i, j),\; m = j)$ declaring that the first coordinate $k$ is the antidiagonal index $i + j$ and the second coordinate $m$ is the position within that antidiagonal.

FILTR provides built-in names for coordinate functions common in bioinformatics: \textit{antidiag(i,j)} groups cells by antidiagonal index, \textit{diag(i,j)} groups cells by diagonal index. Users may also define custom linear combinations, though antidiagonals and diagonals cover the common cases in two-dimensional sequence alignment, sequence prediction, and RNA folding.

\paragraph{RIR Rewrite}

The shearing transformation rewrites the RIR from the original coordinate system into the sheared coordinates $(i, j) \mapsto (k, m)$. For edit distance, the RIR before and after the transformation is:

\noindent
\textbf{RIR Before:}
\[
F_{i,j} =
\min\!\big(
F_{i-1,j-1} + [\mathrm{q}_i \neq \mathrm{r}_j],\;
F_{i-1,j} + 1,\;
F_{i,j-1} + 1
\big)
\]
\[
0 \le i \le M,\qquad 0 \le j \le N
\]

\noindent
\textbf{RIR After:}
\[
F_{k,m} =
\min\!\big(
F_{k-2,m-1} + [\mathrm{q}_{k-m} \neq \mathrm{r}_m],\;
F_{k-1,m} + 1,\;
F_{k-1,m-1} + 1
\big)
\]
\[
0 \leq k \leq M + N,\qquad
\max(0, k - M) \leq m \leq  \min(k, N)
\]

The values computed are identical, but the recurrences expression has completely changed its indexing. Every occurrence of $i$ has been replaced by $k - m$ and every occurrence of $j$ by $m$. The dependency geometry has changed. In the original RIR, dependencies point to $(i-1, j-1)$, $(i-1, j)$, and $(i, j-1)$, a mix of row and column neighbors. In the rewritten RIR, all dependencies point to antidiagonals $k-1$ or $k-2$. DP cells within the same antidiagonal $k$ are independent. The domain has also changed shape: the original rectangle has become a parallelogram, with inner bounds on $m$ that depend on the outer variable $k$.

Two features of the RIR make this transformation possible. First, the RIR represents domains as symbolic bounds that may depend on outer loop variables, so the non-rectangular iteration space that results from shearing---where the range of $m$ changes at each antidiagonal $k$---can be expressed. Second, the RIR treats index variables as symbols within recurrence expressions, so the compiler can rewrite them by substitution without knowing what the recurrence computes.

At a high level, the rewrite introduces a new index variable $k = \textit{antidiag}(i,j)$ representing the antidiagonal, substitutes the original indices with expressions in the new coordinates, and adjusts the loop bounds to reflect the reshaped iteration space. Because shearing is an invertible coordinate change, the compiler can compute an inverse mapping from new coordinates to old. For
$\textit{shear}(k = antidiag(i,j), m = j)$, the inverse is $i = k - m, j = m.$ 

The original domain is a rectangle:
$0 \le i \leq M, 0 \le j \leq N.$ Substituting the inverse mapping yields constraints on the new variables: $0 \le k - m \leq M, 0 \le m \leq N.$ These constraints define a parallelogram in $(k, m)$ space, and the compiler must express it as nested loop bounds in loop order. For $\textit{loopOrd}(k, m)$, the compiler first solves for the outer variable $k$ by eliminating $m$, then solves for the inner variable $m$ given $k$.

Shearing is valid when it defines an invertible change of coordinates and maintains a one-to-one correspondence between original and transformed loop indices. The compiler verifies validity by computing the inverse mapping and checking that the transformed constraints can be expressed as nested loop bounds in the chosen loop order.

\subsection{Value-based Iteration Through Search}
\label{sec:search}

\begin{wrapfigure}{r}{0.45\textwidth}
  \centering
  \includegraphics[width=1\linewidth]{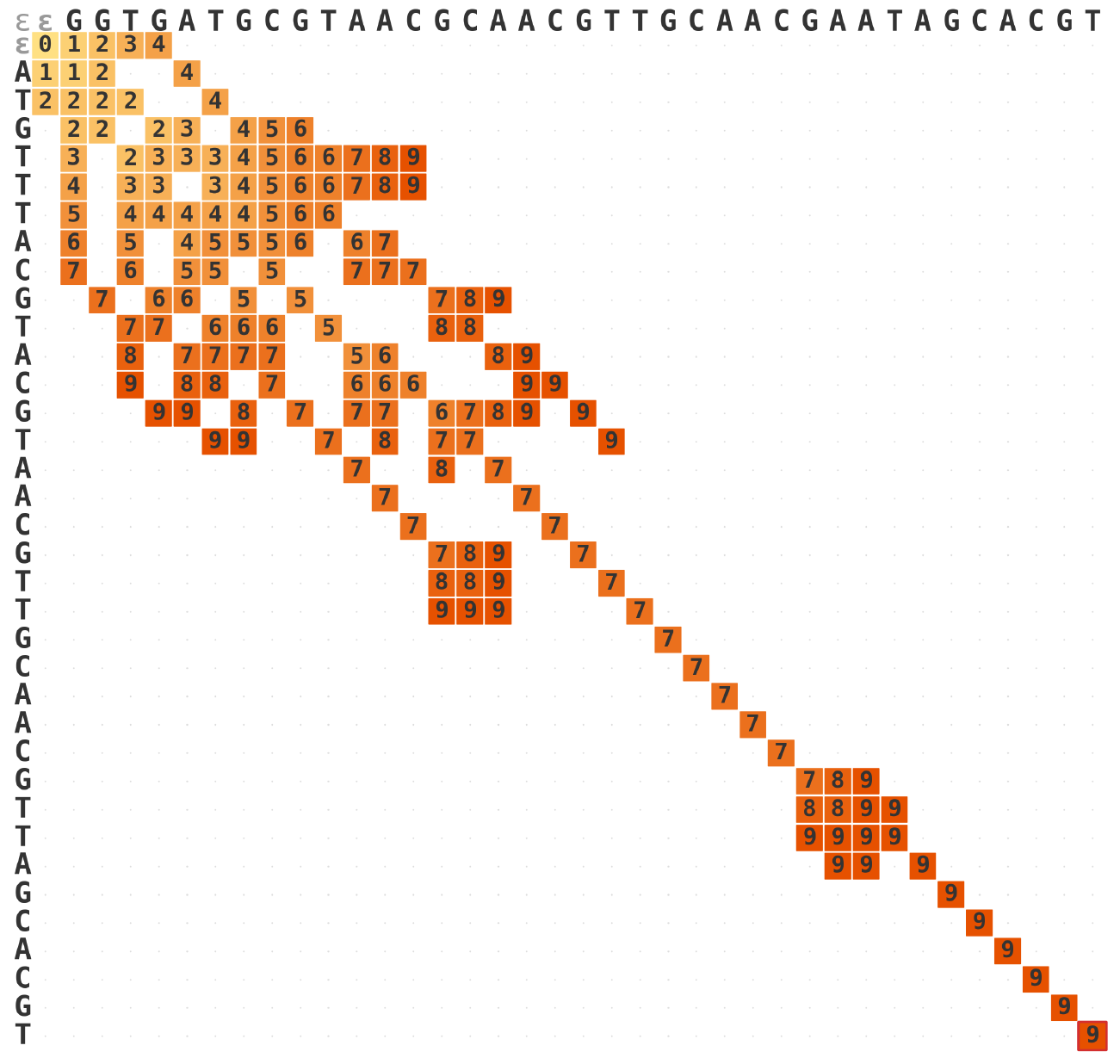}
  \caption{Search iteration for edit distance. Each cell is colored by the number of edits needed to reach it.}
  \label{fig:search-transformation}
\end{wrapfigure}

Loop reordering and shearing both reorganize how cells are visited and stored, but they still visit every cell in the matrix. For edit distance between two sequences of length $M$ and $N$, that means filling all $M \times N$ entries---even when the two sequences are nearly identical and the answer turns out to be a small number of edits. If two genomes of length three billion differ by only a few million edits, the final score is a tiny number relative to the matrix dimensions. Thus, the vast majority of cells hold large edit distances that play no role in the optimal alignment. Filling them is unnecessary.

As illustrated in \Cref{fig:search-transformation}, there is a more natural way to think about traversal~\cite{myers1999, marcosola2021wfa}. Instead of sweeping through positions and computing costs, the iteration can sweep through costs and discover positions. The search starts with zero edits and sees how far it gets. Then it allows one edit and sees where that takes it, then two edits, and so on. Once the search has reached the end of the matrix, it has found the edit distance (which is very small for similar sequences like the sequences in~\Cref{fig:search-transformation}).

The transformation applies to recurrences that find an optimal alignment or path by minimizing or maximizing a score, because these are fundamentally shortest-path problems: each possible alignment corresponds to a path through the DP matrix. Therefore, the recurrence can be reformulated to find the path with the lowest (or highest) total cost to the endpoint.

For example, as described in \Cref{sec:filtr-vs-libraries}, Needleman-Wunsch and affine gap alignment both search for an optimal alignment by choosing, at each cell of the DP matrix, the best among matching, mismatching, inserting, or deleting. This same structure holds for many biology problems and in particular applies to recurrences whose aggregator is a $\min$ or $\max$, and in which all costly transitions have strictly positive weights (for $\min$) or strictly negative weights (for $\max$), ensuring that scores change monotonically along any path through the matrix.

\paragraph{Language feature}

The search transformation implements this by inverting the roles of index and value in the recurrence. The standard edit distance recurrence indexes by position---cell $(i, j)$ stores the edit distance between the first $i$ characters of the query and the first $j$ characters of the reference. The user requests the inversion through the \textit{search} command: for example, $\textit{search}(\textit{score},\ \textit{diag}(i, j))$. 

The keyword \textit{score} tells the compiler to use the value computed by the original recurrence---i.e.\ the edit distance---as the new outer index. Instead of sweeping through all positions and discovering costs, the transformed recurrence sweeps through costs and discovers positions. The remaining coordinate, \textit{diag}$(i, j)$, organizes (e.g., indexes) cells within each score level.

The \textit{score} argument is fixed---it is always the value of the original recurrence---but the second argument is a user-specified index variable. The only constraint is that transitions of \emph{cost zero} must preserve whatever indexing variable is chosen. For edit distance, a character match advances both $i$ and $j$ by one. The transition from $F_{i-1,j-1}$ to $F_{i,j}$ does not incur a scoring cost during a match (as per the edit distance recurrence), and the diagonal $d = j - i$ is unchanged. Therefore, \textit{diag}$(i, j)$ satisfies the constraint. The compiler can verify the constraint automatically: it checks that every zero-cost transition in the original recurrence leaves the chosen index variable unchanged.

The derived recurrence is a non-obvious transformation; the intuition for why this works, and how the compiler derives the new recurrence, is described over the course of this section. The result of this transformation is that for similar sequences the algorithm terminates as soon as it reaches the end of both sequences and only explores cells with less cost than the final edit distance as seen in \Cref{fig:search-transformation}.

\paragraph{RIR Rewrite}

The transformation hinges on a distinction that some transitions are \emph{free} and some are \emph{costly}. In edit distance, a diagonal step where the characters match costs nothing---the alignment advances in both sequences without spending an edit. The other transitions---substitution (mismatch), insertion, deletion---each cost exactly one. Any recurrence amenable to this transformation has the same two-part structure.

This partition maps naturally onto a graph perspective. The alignment matrix is analogous to a directed graph in which each cell is a node. Costly transitions---insertions, deletions, mismatches---are weighted edges (cost 1). Free transitions---character matches---are zero-weight edges. The search transformation is then a form of Dijkstra-like exploration on this graph: it processes nodes in order of cost, expanding outward from the origin one score level at a time until the endpoint (endpoint is the bottom-right corner of the original matrix).

But the structure of the edit distance graph allows something more efficient than generic shortest-path search. Free transitions have a nice geometric property: they preserve the $diag(i,j)$ coordinate. A match advances both $i$ and $j$ by one, so $d = j - i$ is the same before and after. Furthermore, the score does not change during a match. In the edit distance equation, this can be observed because the diagonal transition is defined by $F_{i,j} = F_{i-1,j-1}+[q_i\neq r_j]$. During a match, $q_i = r_j$, meaning $F_{i,j} = F_{i-1,j-1}$. This means a run of consecutive matches stays on a single diagonal and preserves the score, so the entire run can be collapsed into one number---the furthest row reached. For example in \Cref{fig:search-transformation}, the long run of 7s and 9s along diagonal 0 corresponds to a run of matches. 

Costly transitions, by contrast, shift the diagonal by a fixed amount. A deletion steps right ($j$ increases), shifting to diagonal $d + 1$. An insertion steps down ($i$ increases), shifting to diagonal $d - 1$. A mismatch steps diagonally on a non-matching pair, staying on diagonal $d$. The complete set of diagonal shifts is $\{-1, 0, +1\}$, and each shift contributes one term to the three-way $\max$ in the transformed recurrence.

\begin{figure}
    \centering
    \includegraphics[width=1\linewidth]{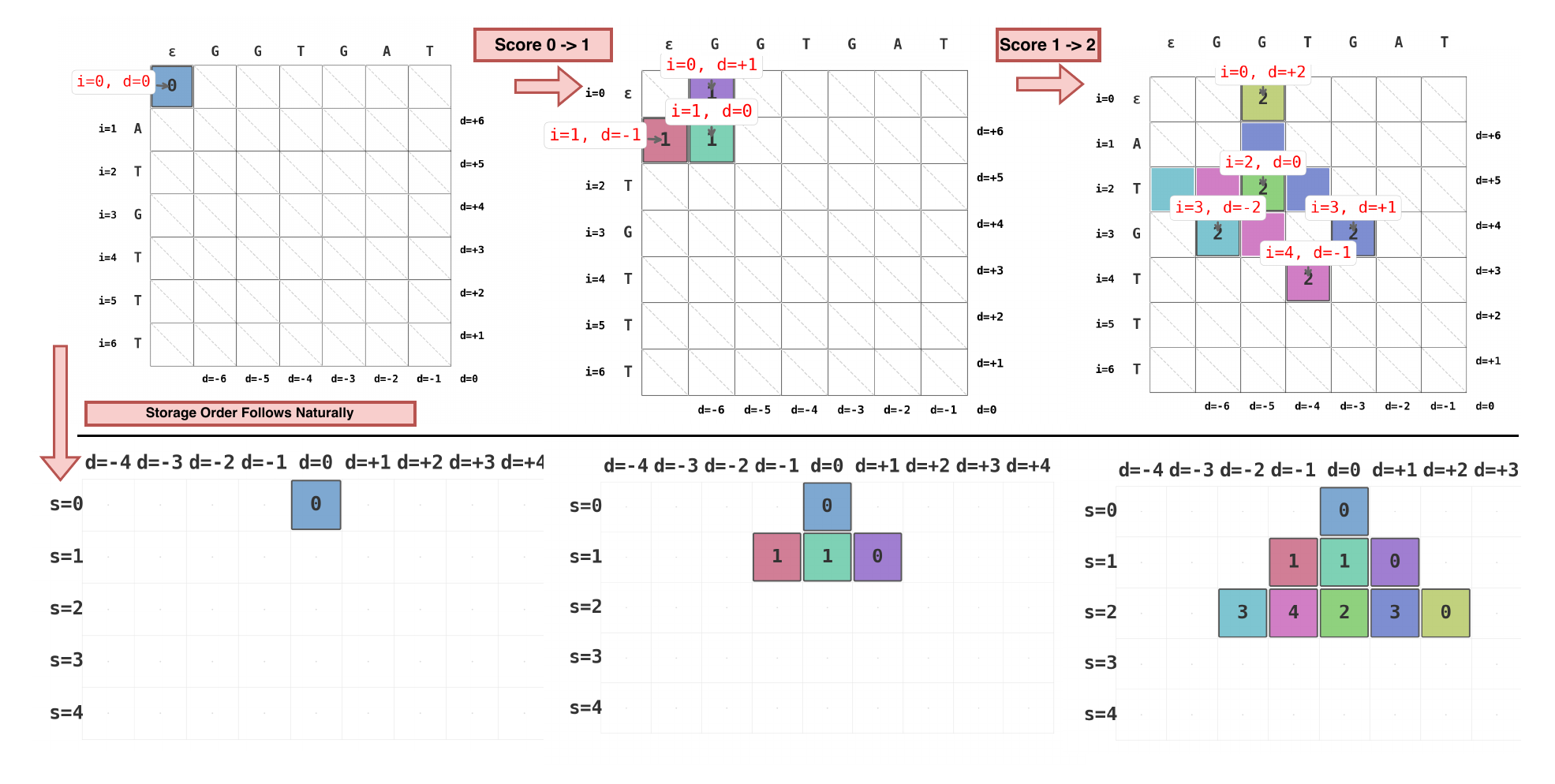}
    \caption{Step-by-step search expansion. Each panel shows one score level's frontier over the original $(i,j)$ matrix (top) and the corresponding row in the cost-indexed recurrence (bottom). The color of each cell on the top corresponds with the color of the cell on the bottom. Notice that the upper matrices store score and the lower ones store indices.}
    \label{fig:wfa-six-diagram}
\end{figure}

\Cref{fig:wfa-six-diagram} walks through the transformation on a concrete pair of sequences. Each panel shows one score level's expansion over the original matrix (top) and the corresponding row in the cost-indexed recurrence (bottom). The storage order (bottom) follows naturally from the iteration pattern (top).

\textbf{Score 0.} The search begins at a single node: the top-left corner of the alignment matrix, sitting on diagonal 0 at row 0. 

\textbf{Score 1.} From every position on the score-0 wavefront, the iteration asks ``what are the neighboring cells or edges that can be reached by spending exactly one edit?'' There are three kinds of moves---a substitution (step diagonally, staying on the same diagonal $d$), an insertion (step down, shifting to diagonal $d-1$), or a deletion (step right, shifting to diagonal $d+1$).  

\textbf{Score 2.} The iteration repeats: each score 1 point generates three edit neighbors, reaching new diagonals at score 2. When two paths arrive at the same diagonal, only the one that reaches further survives---the other is dropped, since it spent the same number of edits but made less progress towards the endpoint. After creating this initial set of neighbors, each surviving DP cell can move forward through consecutive character matches for free. Next, the points on diagonals $-2$, $-1$, and $1$ each advance through matching characters (G$=$G, GT$=$GT, and TG$=$TG) without spending an edit. The purple point at $(d = +1, \text{score } 1)$ is an illustrative example. It fans out to diagonals $0$, $+1$, and $-1$ at score 2 (\Cref{fig:wfa-six-diagram}). The move to $d-1$ is dropped---that diagonal was already reached at score 1 with fewer edits, however the move to $d$ and $d+1$ continue. Move $d$ can advance through the two matching characters (TG$=$TG).

The storage layout and the recurrence definition follow naturally from the search intuition. At each score level, the iteration strategy only touches a small set of diagonals---three at score 1, at most five at score 2, and so on. The iteration never revisits a lower score. So the optimized storage can be just an array per score level, indexed by diagonal, where each entry holds one number: the furthest row reached, since the goal of the algorithm is to reach the last $(\text{row}, \text{col})$. In the original recurrence, the algorithm terminates after filling all $M \times N$ entries; the answer sits in the bottom-right corner $F_{M,N}$. In the transformed recurrence, it terminates as soon as some score $s$ reaches that same corner---i.e.\, when $F_{s,\, N-M} \ge M$. That score is the edit distance.

\noindent \textbf{RIR Before:}
\[
F_{i,j} = \min\!\big( F_{i-1,j-1} + [\mathrm{q}_i \neq \mathrm{r}_j],\; F_{i-1,j} + 1,\; F_{i,j-1} + 1 \big)
\]
\[
0 \leq i \leq M, \qquad 0 \leq j \leq N
\]

\noindent \textbf{RIR After:}
\[
F_{s,d} = \max\!\bigl( F_{s-1,d} + 1,\; F_{s-1,d+1} + 1,\; F_{s-1,d-1} \bigr) + \operatorname{extend}(s,d)
\]
\[
0 \leq s \leq M+N, \qquad \mathrm{diag\_min}_s \leq d \leq \mathrm{diag\_max}_s
\]
\[
\mathrm{diag\_min}_s = \mathrm{diag\_min}_{s-1} - 1, \qquad
\mathrm{diag\_max}_s = \mathrm{diag\_max}_{s-1} + 1
\]

The indices are now score $s$ and diagonal $d$ instead of positions $i$ and $j$. The $\min$ has become a $\max$: the original recurrence minimizes cost, while the new one maximizes position (the furthest row reached on each diagonal). The three terms inside the $\max$ correspond to the three edit operations---mismatch, deletion, insertion---but expressed as shifts between diagonals rather than steps between cells. An $\operatorname{extend}$ function follows the $\max$, greedily advancing through consecutive character matches for free. And the domain is no longer a fixed rectangle; it expands at each score level, with recursive bounds that widen by one diagonal in each direction per score.

Two features of the RIR make this expressible. First, the RIR allows domain bounds that depend on the previous level (domain recurrences) and are computed by auxiliary recurrences, so the expanding diagonal range can be represented directly as recursive bounds. Second, the RIR separates recurrence expressions from domain constraints, allowing the compiler to replace the position-indexed recurrence with a cost-indexed one while independently deriving the expanding domain.

The partition into free and costly transitions of the original recurrence determines the structure. Every free transition---those with zero cost in the original recurrence---goes into the $\operatorname{extend}$ function, since these are the moves that can be followed greedily without increasing the score.

The transitions determine the terms of the new recurrence. Each such transition shifts the non-score coordinate by a fixed amount; the compiler collects these into a shift set. For edit distance, the shifts are $\{-1, 0, +1\}$, producing three terms in the $\max$. A recurrence with different edit operations---say, one that also allowed swapping of adjacent characters---would have a different shift set and a correspondingly different number of terms.

The shift set also determines the expanding domain. At score 0, only the origin diagonal is active. Each transition can shift the diagonal by at most 1, so the range of active diagonals widens by at most 1 in each direction per score level---the recursive domain bounds in the RIR. A recurrence whose transitions could shift by up to $k$ would see its domain widen by $k$ per level instead.

Finally, the termination condition follows from the coordinate mapping: the original recurrence terminates when it fills cell $(M, N)$; the transformed recurrence terminates when diagonal $N - M$ reaches row $M$, which is the same cell expressed in the new coordinates.

\section{Pruning Language}
\label{sec:pruning-rewrites}

\begin{wrapfigure}{r}{0.45\textwidth}
  \vspace{-1em}
  \centering
  \begingroup
  \setlength{\arraycolsep}{2pt}
  \scriptsize
  \begin{minipage}[t]{\linewidth}
\[
\begin{array}{r@{\;}l@{\;}l}
  \textsf{PruneProg} & ::=
    \textsf{PruneConstruct}^{*} \\[0.2em]
  \textsf{PruneConstruct} & ::=
    \textsf{PruneAxis}
    \mid \textsf{Stmt} \quad \textit{(from~\Cref{fig:rir-grammar})} \\[0.2em]
  \textsf{PruneAxis} & ::=
\textsf{Expr} \, \textsf{CmpOp} \, \textsf{CoordExpr} \, \textsf{CmpOp} \, \textsf{Expr} \\[0.2em]
  \textsf{CoordExpr} & ::=
    \textit{diag}(v, v)
    \mid \textit{antidiag}(v, v)
    \mid v | 
    \mid ...
\end{array}
\]
  \end{minipage}
  \endgroup
  \caption{Grammar of the Pruning Language. Pruning-language is syntactically RIR statements but written in user coordinates and compiled into loop order coordinates by the compiler.}
  \label{fig:pruning-language-grammar}
\end{wrapfigure}

In two-dimensional sequence alignment, each dynamic programming recurrence corresponds to an alignment matrix. Matrix cells near the main diagonal represent alignments with relatively few insertions or deletions, while cells far from the diagonal correspond to large deviations between the reference and query sequences. When two sequences are expected to be similar, exploring far-off-diagonal cells is both unnecessary and wasteful. This reflects a basic biological intuition: mutations tend to be rare and spread out, and alignment recurrences should focus computation on regions where the sequences align well. More generally, many bioinformatics recurrences have large regions of the matrix that are unlikely to contribute to the final answer. 

Pruning restricts which parts of the matrix get computed, skipping these regions entirely. Because pruning discards cells that the original recurrence computes based on heuristics, it is a non-semantics-preserving transformation (final recurrence values and traceback path may be different to the unpruned, unscheduled version). The user is therefore responsible for ensuring---based on biological intuition---that the omitted regions do not significantly affect the final result. The grammar for FILTR's pruning language is given in \Cref{fig:pruning-language-grammar}.

The pruning language provides a language feature---the \emph{prune axis}---which is an inequality that restricts iteration to a geometric sub-region of the matrix. For example, a band around the main diagonal is denoted by \textit{prune axis}: $-B \le \mathrm{diag}(i,j) \le B$. The user writes an inequality that bounds a coordinate expression (e.g., $\mathrm{diag}(i,j)$)  over the loop indices, and only cells satisfying that inequality are ever visited. The coordinate expression can be any function of the indices that is invertible, but generally is just $\mathrm{diag}(i,j)$ or $\mathrm{antidiag}(i,j)$. The bounds of a prune axis may also be recurrence expressions rather than constants, allowing the geometric region to evolve dynamically at runtime. Therefore, a pruning program may be either \emph{static}, with constant bounds, or \emph{dynamic}, where the user can specify domain recurrences that update the prune axis bounds at runtime.

The Pruning Language serves two purposes: (1) it separates concerns (pruning, iteration, and base recurrence) for the user, and (2) it allows for a flexible mixing of coordinate systems.~\Cref{fig:pruning-language-x-drop-spec} shows $k$ and $m$ from the antidiagonal coordinate system being used in tandem with $i$ and $j$. Specifically, the recurrences should be indexed by $k$ and $m$ (i.e.\, the \textit{loopOrd} variables) but coordinate expressions (e.g. $\mathrm{diag}(i,j)$) can be defined in  either coordinate system (e.g., $\mathrm{diag}(i,j)$, $j-i$, or $2m-k$).

The pruning model focuses on supporting real-world heuristics but does not allow for arbitrary pruning. Specifically, it targets pruning techniques over contiguous geometric regions, assuming that at each step there is a single continuous active region and that pruning decisions rely only on information available at the current step rather than future values. In practice, we have not encountered algorithms that violate either assumption, and the language can be easily extended to support additional pruning needs in the future.

Although pruning programs are often longer than those in the Recurrence Language or the Iteration Ordering Language, each one is still small (\Cref{fig:pruning-language-x-drop-spec} shows the five line X-drop pruning program) and can be applied to many different recurrence equations. This same X-drop pruning program, for example, can be composed with edit distance, affine gap alignment, or other recurrences as described in \Cref{sec:evaluation}.

\subsection{Static Pruning}
\label{sec:static-pruning}
In static pruning, the bounds are fixed at compile time before execution. For example, in banded edit
distance, if the user knows in advance that the best alignment will stay
within $B$ diagonals of the main diagonal, then they can use static pruning. The pruning program is a single
\textit{prune axis}: $-B \;\le\; \mathrm{diag}(i,j) \;\le\; B$.

The user can write this constraint in whichever coordinate system is most natural, as long as the expression can be algebraically rewritten in  terms of the \textit{loopOrd} variables. To illustrate, the compiler internally knows that $diag(i,j) = j-i$ and $antidiag(i,j)=i+j$. FILTR automatically converts the user-provided constraints so they are defined strictly in terms of the desired loop iteration variables (using the sympy computer algebra system). The compiler then intersects the rewritten constraint bounds with the existing domain bounds. For col-major traversal with $\textit{loopOrd}(j,i)$, the compiler solves $\mathrm{diag}(i,j) = j - i$ for $i$, yielding
$0 \le j \le N,\; \max(0, j{-}B) \le i \le \min(M, j{+}B)$. Under
$\textit{loopOrd}(k, m)$ where $k=antidiag(i,j)$ and $m=j$, the compiler algebraically solves $\mathrm{diag}(i,j) = j-i$ for $m$ and $k$, which resolves to  $\mathrm{diag}(i,j)=2m-k$, yielding
$0\le k\le M+N,\quad
\max\bigl(0,\,k-M,\,\lceil (k-B)/2\rceil\bigr)\le m\le
\min\bigl(N,\,k,\,\lfloor (k+B)/2\rfloor\bigr)$. Under a \textit{search(score, diag(i,j))} transformation, the constraint maps to $-B \le d \le B$ in the $(s, d)$ loop space.

\subsection{Dynamic Pruning}
\label{sec:dynamic-pruning}
To compute dynamic prune axis bounds at runtime, the Pruning Language provides three additional constructs---recurrence equations, conditional recurrences, and set-builder reductions---that are syntactically a subset of the RIR but written with respect to the Recurrence Language input. These equations allow the user to describe the domain as a recurrence (domain recurrence), track auxiliary information such as a running minimum score, and conditionally override a cell's value (e.g., replacing it with $\infty$ when a pruning condition is met).

\begin{wrapfigure}{r}{0.55\textwidth}
  \centering
    \vspace{-1em}
  \includegraphics[width=1\linewidth]{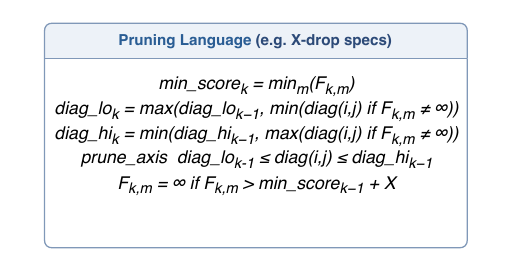}
    \vspace{-2em}
   \caption{Pruning specification for X-drop.}
  \label{fig:pruning-language-x-drop-spec}
  \vspace{-1em}
\end{wrapfigure}

In \emph{dynamic pruning}, the boundaries adapt at runtime based on computed values. This requires all the constructs working together, as shown in X-drop in \Cref{fig:pruning-language-x-drop-spec}, an auxiliary recurrence analyzes the most recently computed cells ($min\_score_{k}=$), a conditional recurrence uses this analysis to mark cells for removal ($F_{k,m} = \infty \;\text{if }$), set-builder reductions calculate the new bounds based on these markings ($\min(\text{diag}(i,j)) \;\text{if } F_{k,m} \neq \infty$). The result is a feedback loop---what is computed (data recurrence) determines where computation happens next (domain recurrences).

\setlength{\columnsep}{0pt}

Other pruning programs follow a similar structure. Z-drop, for instance, replaces the X-drop conditional with one that measures distance in addition to score and is described in \Cref{sec:filtr-vs-libraries}. Similarly, after applying a search transformation to a recurrence, the user can derive a pruned version of the resulting search algorithm, such as WFA-Adapt, which prunes diagonals based on distance from the diagonal closest to the target endpoint (also described in \Cref{sec:filtr-vs-libraries}). 

The compiler lowers this pruning program into the Recurrence IR by converting all expressions into the loop order coordinate scheme, introducing $\mathrm{min\_score}_k$ as a recurrence, splitting the main recurrence into conditional regions, and substituting the domain recurrences for static loop bounds. The resulting RIR is the X-drop formulation shown in \Cref{fig:six-panel-RIR}~(right). The domain bounds $\mathrm{diag\_lo}_k$ depend on $F_{k,m}$, while the iteration of $F_{k,m}$ is controlled by those same bounds at the previous timestep---this bidirectional dependence between what is computed and where computation happens is what distinguishes dynamic pruning from static banding.

\section{Code Generation}
\label{sec:code-generation}

\begin{figure}
    \centering
    \includegraphics[width=1\linewidth]{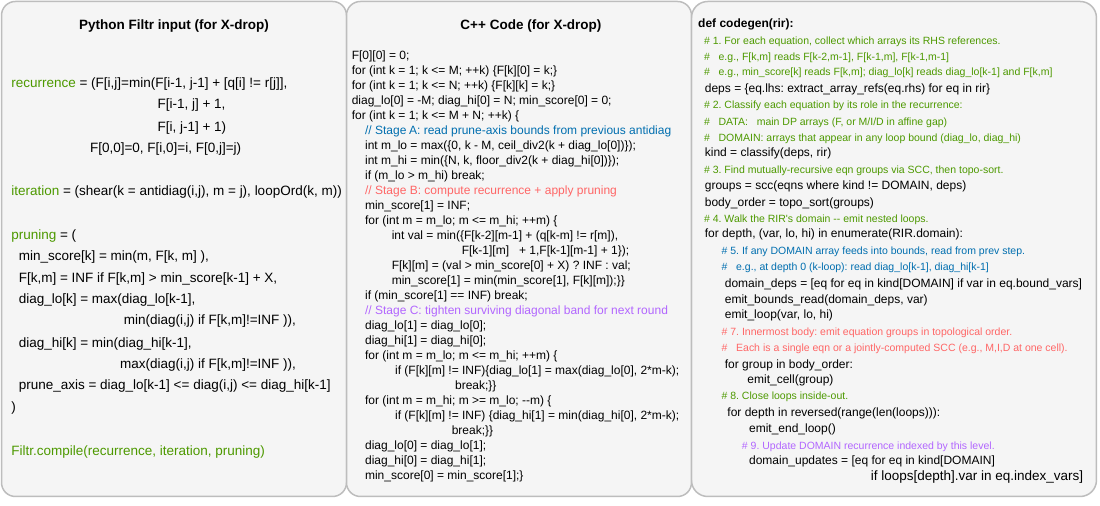}
    \caption{End-to-end example of FILTR compiling X-drop edit distance, showing the user-facing input specification (left), simplified generated C++ (center), and the code generation algorithm (right). The C++ code is generated code slightly cleaned for readability.}
    \label{fig:code-gen}
\end{figure}

FILTR's compiler transforms the rewritten RIR into C++ code. We adopt the general approach of the Recuma recurrence compiler~\cite{recuma}, which constructs a dependency graph from recurrence definitions and lowers it into loop code. In our case, the RIR serves as the input to this lowering. \Cref{fig:code-gen} shows the complete pipeline for X-drop edit distance: the user-facing FILTR input (left), the generated C++ (center), and the code generation algorithm (right).

The FILTR compiler first applies iteration rewrites and then applies pruning rewrites. Applying iteration rewrites first allows the compiler to verify that the transformed traversal order respects the recurrence’s data dependencies against the full, unpruned recurrence. Applying pruning second is necessary because dynamic pruning strategies like X-drop decide what to prune based on what has been computed so far, meaning their behavior depends on the order in which states are visited. As such, pruning recurrences must be expressed in the coordinate system produced by the iteration rewrites. For example, an X-drop specification references antidiagonal indices, which only exist after the shearing transformation has been applied.

The main challenge of lowering the RIR to C++ is the dependency between data and domain recurrences. Domain variables such as \text{diag\_lo} and \text{diag\_hi} are defined by recurrences that read from the main DP array, yet those same variables determine the loop bounds over which the DP array is computed. FILTR resolves this through staged code generation within a single loop body, organized into three stages. As shown in the codegen algorithm in \Cref{fig:code-gen} (right), the compiler first classifies each equation by its role: \textit{data} recurrences are the main DP arrays (e.g., $F$, or $M$, $I$, $D$ in affine gap), and \textit{domain} recurrences are arrays that appear in loop bounds (e.g., \text{diag\_lo}, \text{diag\_hi}). It then topologically sorts the non-domain equations to determine the order in which they are emitted. A domain recurrence is emitted immediately after all of its data dependencies are satisfied (i.e.\, after the respective data recurrences that it depends on have been emitted).

The final output arranges arrays according to the iteration order's indexing scheme: for antidiagonal traversal, arrays are indexed by antidiagonal number and offset; for score-based indexing, by score layer and diagonal. FILTR relies on vectorization hints and autovectorization to generate performant C++ code.

\section{Evaluation}
\label{sec:evaluation}

Our evaluation supports three claims: (1) FILTR-generated code matches or exceeds the performance of hand-optimized bioinformatics libraries and existing recurrence compilers; (2) FILTR's composable interface lets users explore combinations of pruning and traversal strategies to find the best configuration for a given dataset; and (3) pruning specifications written for one recurrence transfer directly to others, making it easy to construct new heuristics.

\subsection{Experimental Methodology}
\label{sec:experimental-methodology}
All experiments were conducted on an Intel Core i9-285K workstation with 64\,GiB of RAM. FILTR-generated C++ kernels were compiled with full optimization (\textit{-O3}) and automatic vectorization enabled. Each benchmark was executed ten times; the two highest and two lowest measurements were discarded to reduce the influence of outliers, and the reported runtime for an experiment is the mean of the remaining six runs.

There are four broad classes of bioinformatics algorithms that rely on recurrences: sequence alignment, RNA folding (predicting how an RNA molecule folds into a three-dimensional structure), sequence prediction models (used to identify patterns such as genes within a sequence), and sequence chaining (linking short matching alignments into a longer alignment). There is a diverse set of recurrences for each problem area, but they generally look similar to those in \Cref{tab:dp_recurrences}. Although these recurrences have more complicated dependency structures than edit distance, we can still apply the same diagonal, antidiagonal, and search transformations.

\begin{figure}[t]
  \centering
  \begin{minipage}{0.5\textwidth}
    \centering
    \includegraphics[width=\linewidth]{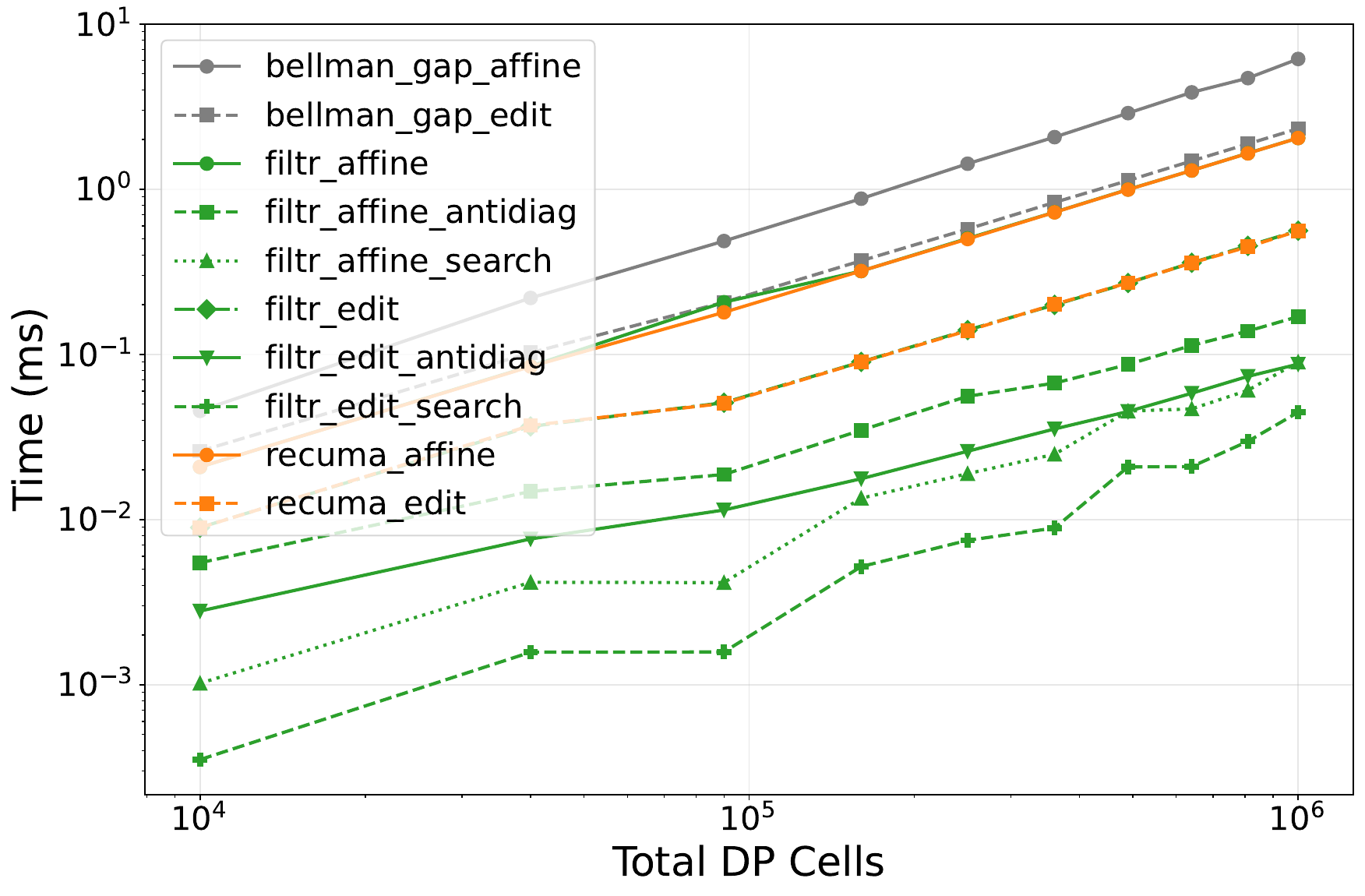}
  \end{minipage}\hfill
  \begin{minipage}{0.5\textwidth}
    \centering
    \includegraphics[width=\linewidth]{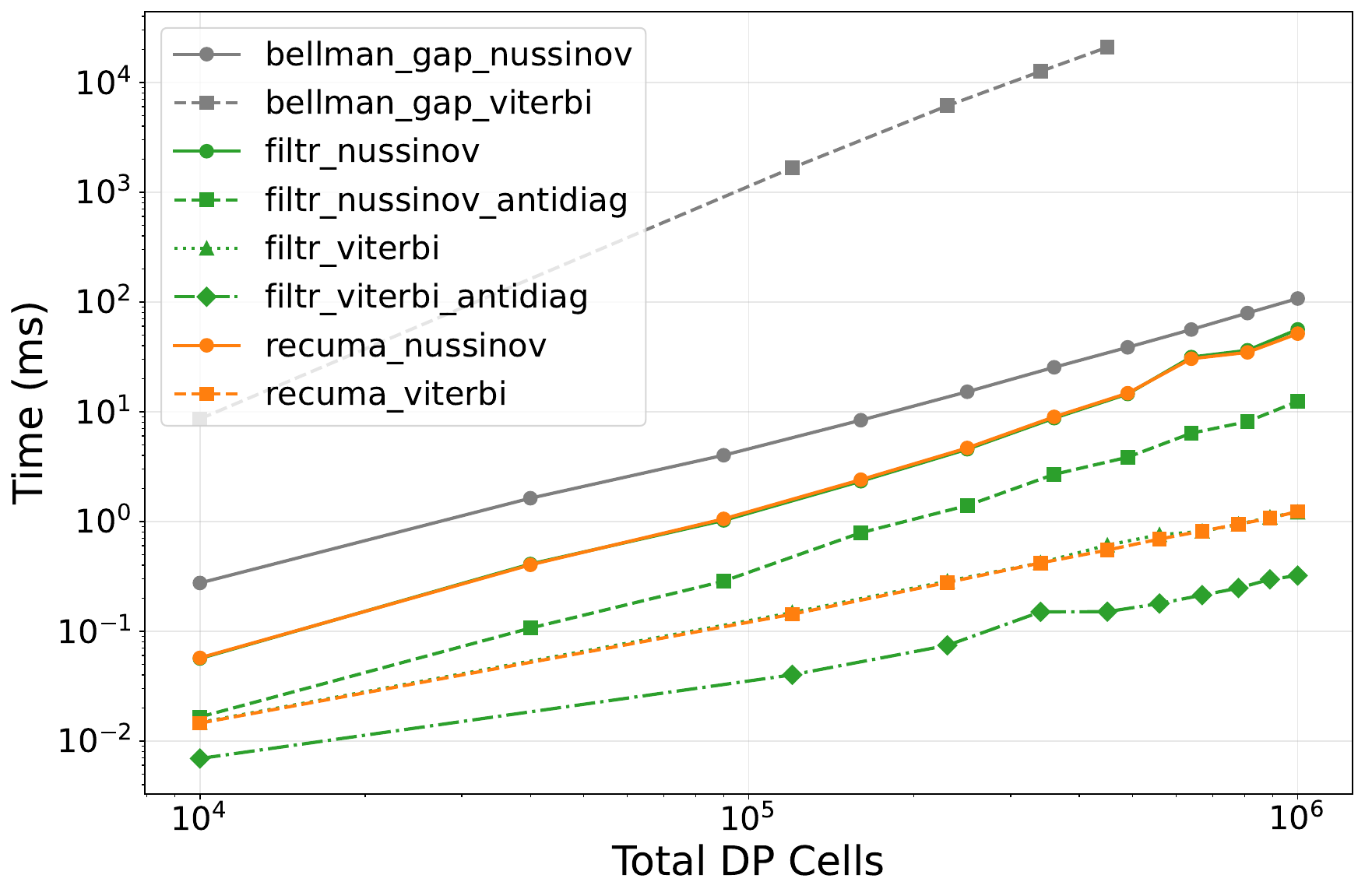}
  \end{minipage}
  \caption{FILTR versus Recuma runtime (log-log) across four recurrences: edit distance, affine gap penalty (left), Nussinov RNA folding, and Viterbi Sequence Prediction (right).}
  \label{fig:recuma-vs-filtr}
\end{figure}

We report runtime in milliseconds and, where applicable, alignment accuracy as the percentage difference between exact and approximate results. Library and compiler comparisons use synthetic sequences so that divergence level, sequence length, and mutation profile can be controlled independently, isolating performance differences. The evaluation of design-space exploration (\Cref{sec:exploring-design-space}) and heuristic transfer (\Cref{sec:transferring-heuristics}) uses a wide variety of real DNA and RNA sequences to assess how FILTR behaves under authentic biological conditions.

\subsection{FILTR vs.\ Recurrence Compilers}
\label{sec:filtr-vs-compilers}
Recuma, the primary compiler baseline, represents the current state of the art in recurrence compilation. It generates efficient code for standard row-by-row and column-by-column traversals of the DP matrix but lacks support for alternative traversal orders, pruning strategies, or score-indexed traversal. Our evaluation datasets include synthetic query--reference sequence pairs that are 90\% similar and representative of many real genomic datasets~\cite{jain2018high}. 

\Cref{fig:recuma-vs-filtr} plots the runtime against the problem size for different configurations across four recurrences: edit distance, affine gap penalty, Nussinov~\cite{nussinov1978} (RNA folding), and Viterbi (sequence prediction). On a standard row-wise traversal, FILTR matches Recuma's runtime closely at all tested sizes, confirming that FILTR's code generation introduces no overhead relative to an existing recurrence compiler. Antidiagonal iteration, which exposes automatic SIMD parallelism, yields a $7$--$13\times$ speedup over Recuma at medium to large sizes for affine gap penalty, and a $6$--$8\times$ speedup for edit distance on asymmetric problems. With the search transformation applied, edit distance achieves $14$--$20\times$ speedups and affine gap achieves $25$--$50\times$. However, as we show in~\Cref{sec:exploring-design-space}, these gains depend on the input data. For example, search performs best on similar sequences and its advantage diminishes as divergence increases. We also compare against Bellman's Gap, which performs comparably to Recuma; however, it is similarly slower than FILTR kernels.

\subsection{FILTR vs.\ Hand-Optimized Libraries}
\label{sec:filtr-vs-libraries}
Sequence alignment is one of the most performance-critical operations in bioinformatics, and the community has invested heavily in hand-tuned, library-level implementations~\cite{li2018minimap2, seqan, marcosola2021wfa, daily2016parasail}. We compare FILTR against four widely used libraries: SeqAn (which provides unpruned, banded, and X-drop alignment), Parasail (which provides unpruned and banded), Ksw2 (which provides an X-drop variant), and WFA2 (which provides a search variant). Parasail, WFA2, and Ksw2 are hand-vectorized implementations.

\begin{figure}[t]
\centering
\begin{minipage}{0.49\textwidth}
  \centering
  \includegraphics[width=\linewidth]{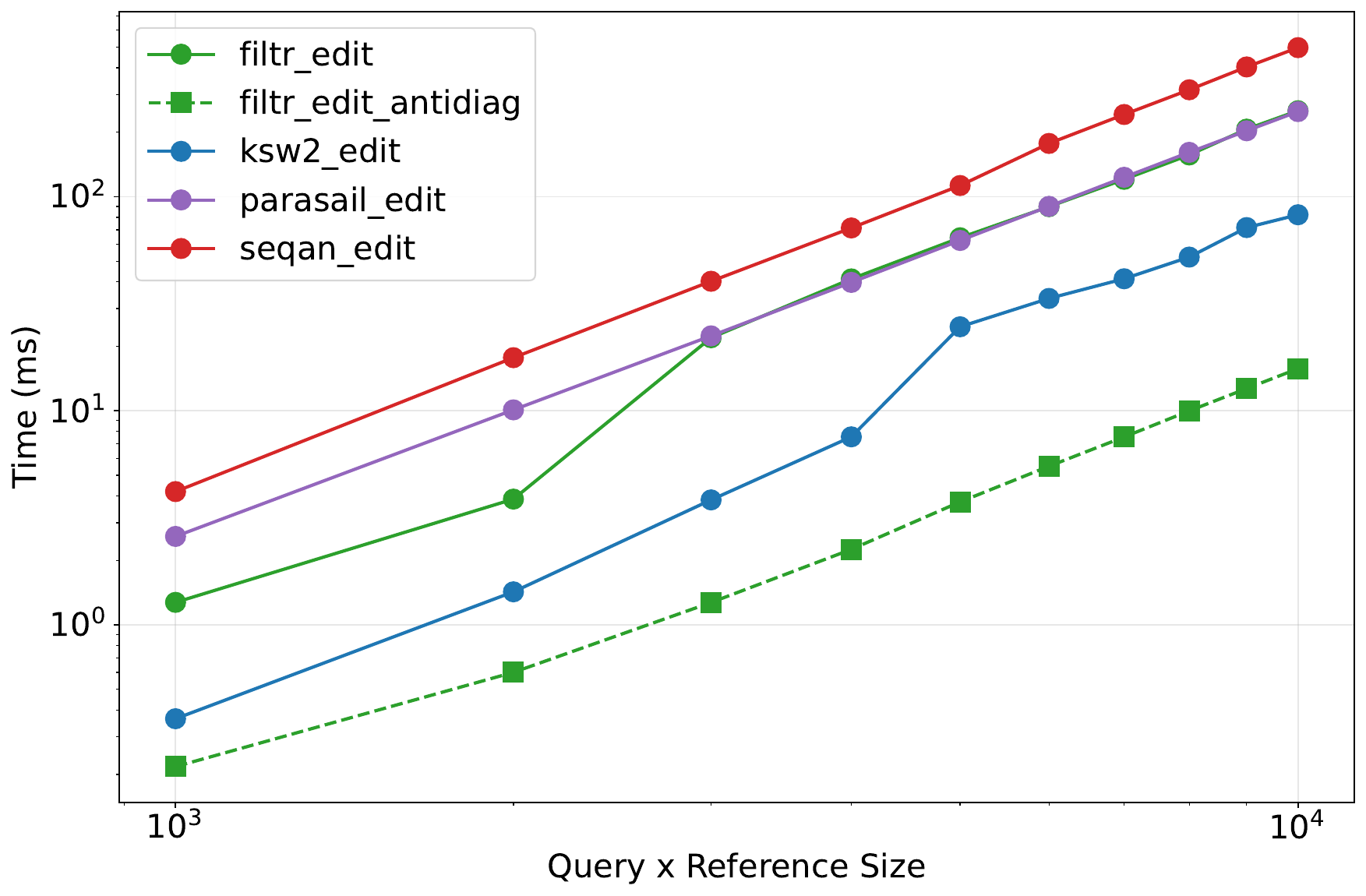}
  \caption{FILTR versus hand-optimized unpruned alignment libraries (Ksw2, Parasail, SeqAn) on edit distance. Log scale.}
  \label{fig:filtr-vs-exact}
\end{minipage}
\hfill
\begin{minipage}{0.49\textwidth}
  \centering
  \includegraphics[width=\linewidth]{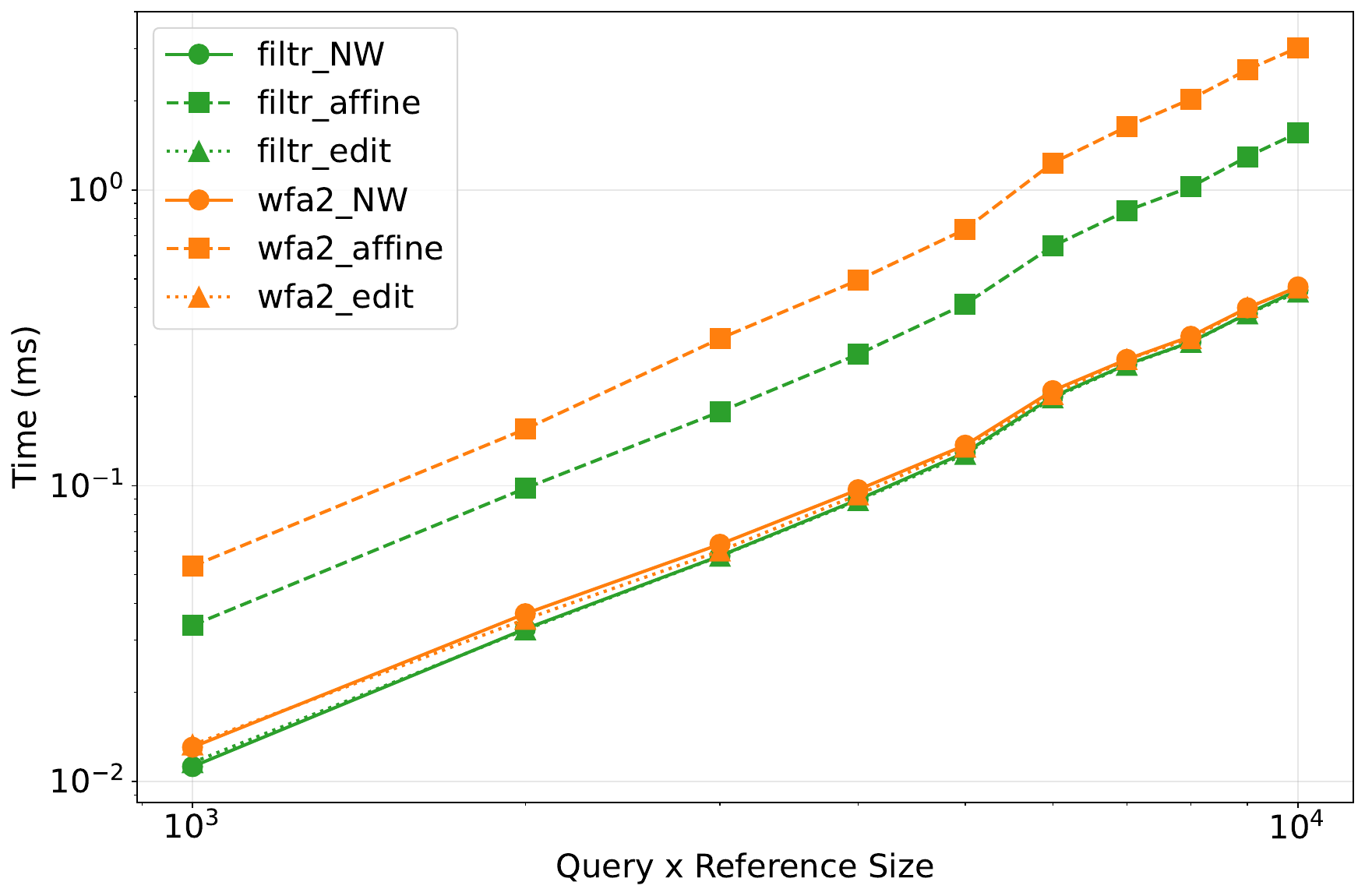}
  \caption{FILTR search vs.\ WFA2. We can additionally apply this search to longest common subsequence, dual gap, chaining, and RNA folding. Log scale.}
  \label{fig:filtr-vs-wfa}
\end{minipage}
\end{figure}

\paragraph{Unpruned comparison.}

The Needleman–Wunsch algorithm and its simpler Edit Distance variant use a recurrence to compute the optimal global alignment between two sequences. As shown in \Cref{fig:filtr-vs-exact}, FILTR antidiagonal iteration is consistently the fastest implementation. The baseline methods are $5$--$30\times$ slower than FILTR's antidiagonal shearing. This is because their vectorized implementations do not use a cache-efficient memory layout, whereas FILTR's shearing iteration transformations ensure that consecutively traversed cells are contiguous in memory.

We also compare FILTR's search transformation against the hand-optimized WFA2 library since this is another exact method. The search transformation corresponds to the wavefront algorithm (realized in WFA2~\cite{marcosola2021wfa}). \Cref{fig:filtr-vs-wfa} compares FILTR and WFA2 across three scoring models---Needleman Wunsch, edit distance, and affine gap penalty. FILTR remains the best. Because FILTR formulates the wavefront traversal as a standard recurrence over score coordinates, the compiler can apply automatic vectorization.

\paragraph{Static pruning.}

\begin{figure}[t]
\centering
\begin{minipage}{0.49\textwidth}
  \centering
  \includegraphics[width=\linewidth]{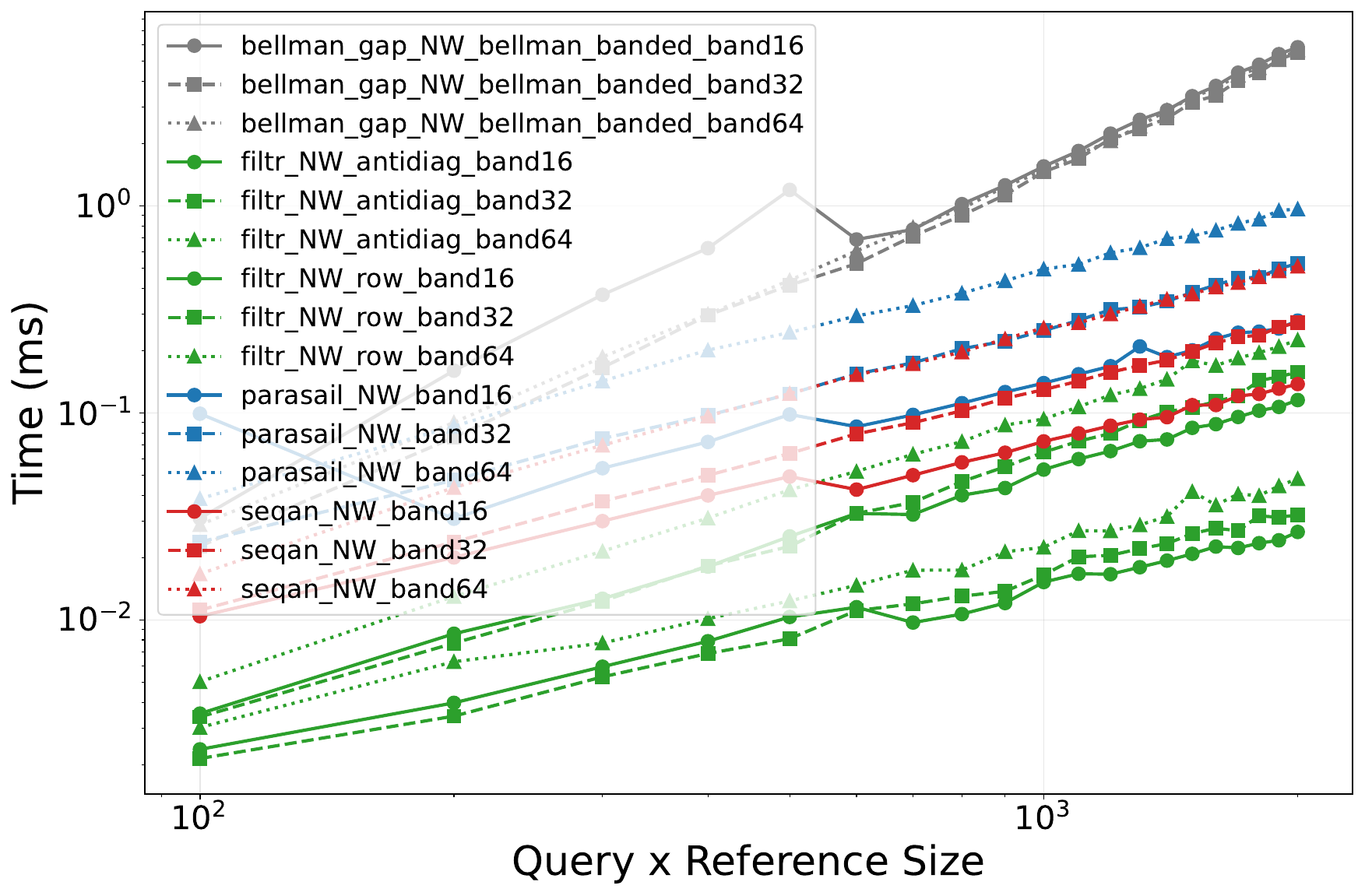}
  \caption{FILTR vs.\ banded alignment libraries and compilers. Log scale. Spike is due to a cache line boundary.}
  \label{fig:filtr-vs-banded}
\end{minipage}
\hfill
\begin{minipage}{0.49\textwidth}
  \centering
  \includegraphics[width=\linewidth]{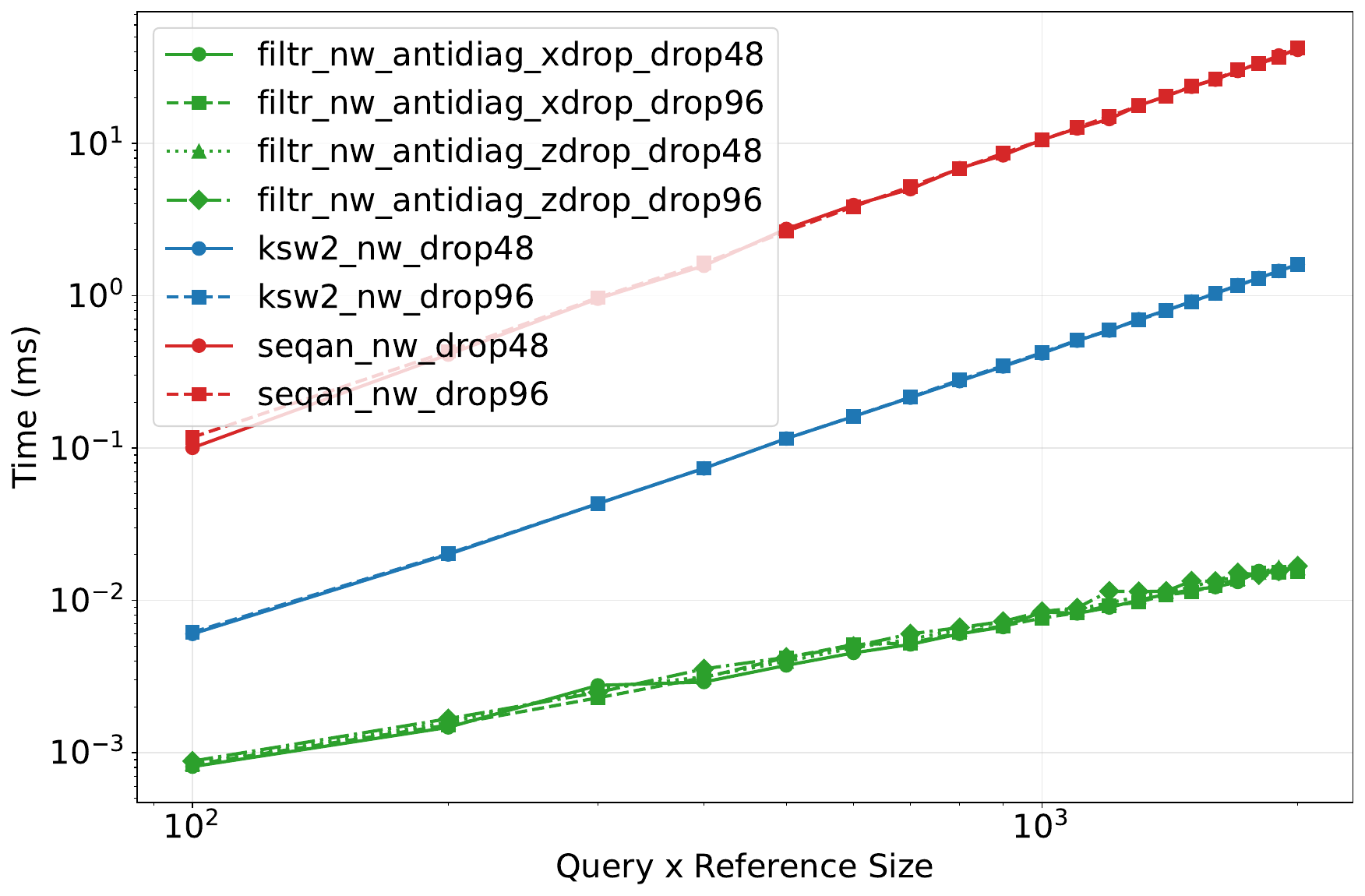}
  \caption{FILTR vs.\ dynamic pruning libraries. Log scale.}
  \label{fig:filtr-vs-dynamic}
\end{minipage}
\end{figure}

FILTR's banded Needleman--Wunsch kernel substantially outperforms SeqAn and Parasail across all tested band sizes (16, 32, and 64). \Cref{fig:filtr-vs-banded} shows that FILTR's antidiagonal banded kernels (green) consistently outperform both Parasail (blue) and SeqAn (red) across all band widths and sequence lengths, often by an order of magnitude at larger sizes. FILTR also outperforms the Bellman's Gap language. FILTR is faster because it applies a shearing transformation to memory layout, enabling efficient access along antidiagonals, whereas other libraries use a hard-coded row-major layout. FILTR's row-wise banded kernels (green) are competitive with or faster than the hand-tuned libraries.

\paragraph{Dynamic pruning.}

Next, we compare heuristic methods that prune computed cells dynamically based on alignment scores rather than a fixed band based boundary. X-drop stops extending the alignment when the current score drops more than a fixed threshold below the best score seen so far. Z-drop is similar but additionally penalizes distance from the best-scoring diagonal, making it more tolerant of extended gaps and score fluctuations that arise from long insertions or deletions. In our language, the conditional recurrence in the pruning program would be: 
\[
F_{k,m} = -\infty \quad \text{if} \quad \mathrm{max\_score}_{k-1} - val > Z + \left| \mathrm{diag}(i,j) - \mathrm{diag\_best}_{k-1} \right|\]

This condition corresponds to the affine-gap Z-drop heuristic. Since affine gap is a max recurrence, cells that violate the Z-drop condition are invalidated by assigning them a score of $-\infty$. Here, $\mathrm{max\_score}$ denotes the best alignment score, at coordinate $\mathrm{diag\_best_{k}}$ (which is determined with \text{argmax} data recurrences). \Cref{fig:filtr-vs-dynamic} shows that FILTR's X-drop and Z-drop variants (green) outperform both Ksw2 (blue) and SeqAn (red) by one to two orders of magnitude across all sequence lengths. Both Ksw2 and SeqAn traverse without shearing the storage structure. Notably, Ksw2 does not provide an X-drop algorithm, and neither SeqAn nor Parasail provides a Z-drop algorithm, whereas FILTR generates both X-drop and Z-drop from the same recurrence specification with different pruning conditions.

\setlength{\columnsep}{0.5em} 
\begin{wrapfigure}{r}{0.5\textwidth}
  \begin{center}
      \vspace{-2em}
    \includegraphics[width=0.5\textwidth]{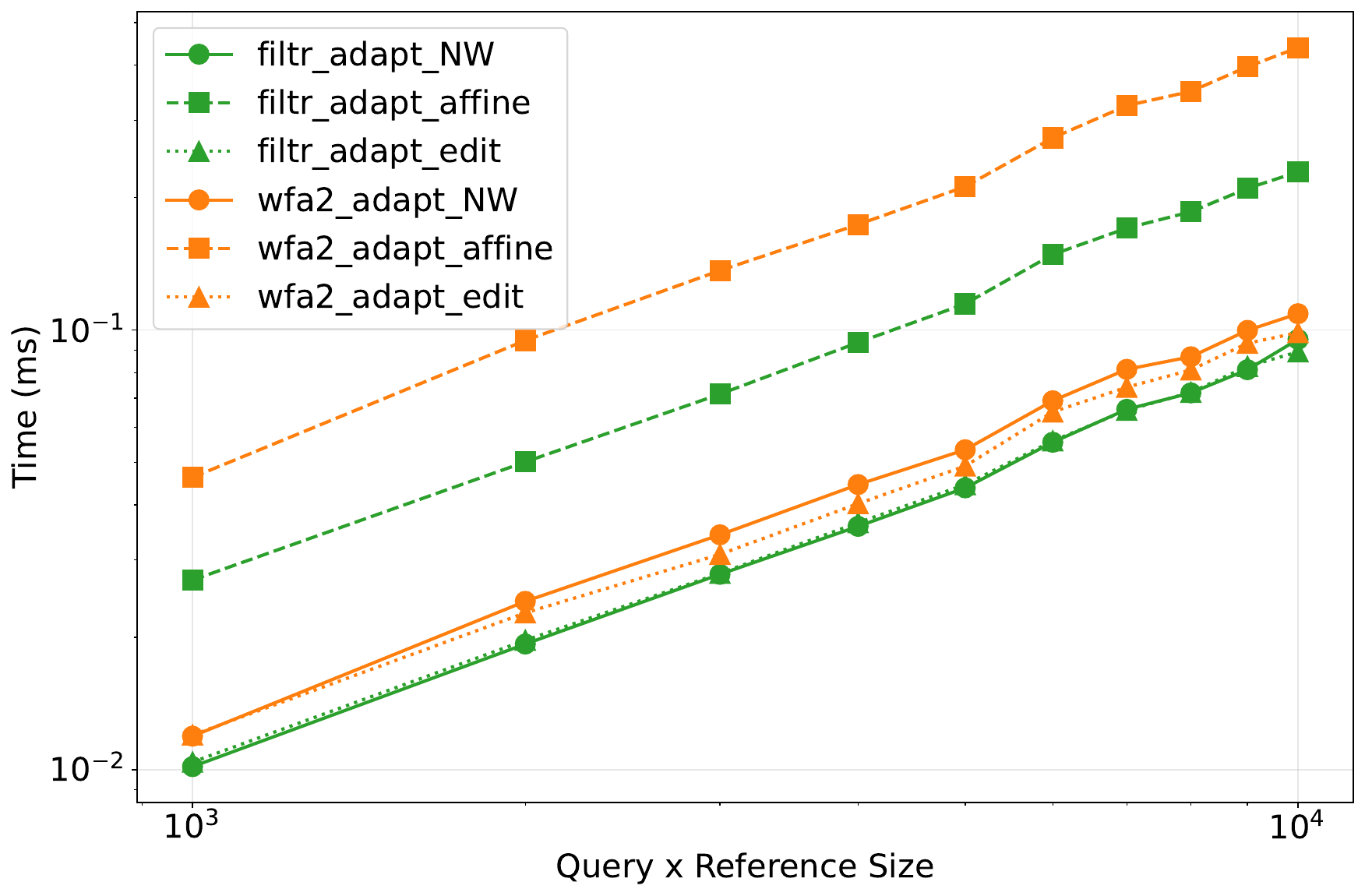}
        \vspace{-1em}
  \end{center}
  \caption{FILTR vs. WFA-Adapt. Log scale.
}
    \vspace{-2em}

  \label{fig:wfa-heuristic}
\end{wrapfigure}

We also compare against dynamic pruning strategies applied to search. The WFA-Adapt algorithm \cite{marcosola2021wfa} employs a heuristic that, at each score level, discards diagonals that fall too far from the leading diagonal (the diagonal representing the most progress through the alignment matrix). FILTR can express this data-dependent pruning and achieves performance comparable to this hand-optimized library implementation (\Cref{fig:wfa-heuristic}). 

\subsection{Exploring the Design Space}
\label{sec:exploring-design-space}
\begin{wraptable}{r}{0.55\linewidth}
  \centering
  \begin{tabular}{ccrrr}
    \toprule
    Div. \% & Size & SR (s) & AD (s) & Speedup \\
    \midrule
    1  & 500$\times$500   & 0.0007 & 0.069  & 97$\times$ \\
    1  & 2000$\times$2000 & 0.0089 & 0.87   & 98$\times$ \\
    1  & 4000$\times$4000 & 0.0138 & 4.15   & 300$\times$ \\
    \midrule
    5  & 500$\times$500   & 0.0019 & 0.050  & 27$\times$ \\
    5  & 2000$\times$2000 & 0.052  & 0.71   & 14$\times$ \\
    5  & 4000$\times$4000 & 0.19   & 4.27   & 23$\times$ \\
    \midrule
    30 & 500$\times$500   & 0.11   & 0.051  & 0.45$\times$ \\
    30 & 2000$\times$2000 & 1.58   & 0.72   & 0.45$\times$ \\
    30 & 4000$\times$4000 & 6.02   & 4.36   & 0.72$\times$ \\
    \bottomrule
  \end{tabular}
    \caption{Speedup of search traversal over antidiagonal traversal for edit distance. SR is search traversal and AD is antidiagonal traversal. (s) denotes seconds.}
        \vspace{-1em}

      \label{tab:search-vs-antidiag}
\end{wraptable}
We have established that FILTR matches or exceeds existing systems on individual benchmarks. Now, we demonstrate that the best heuristic and schedule depend on properties of the input data. This means there is no single best configuration---users need the ability to explore alternative heuristics.

\paragraph{Search vs.\ antidiagonal traversal.}
\Cref{tab:search-vs-antidiag} reports the speedup of search-based (score-indexed) traversal over antidiagonal traversal for edit distance at three divergence levels and three matrix sizes. At 1\% divergence, the search transformation is 97--300$\times$ faster because it searches only the narrow strip of cells at each score level, while antidiagonal traversal fills the entire matrix. At 5\% divergence, the advantage narrows to 14--27$\times$, and at 30\% divergence, search is slower (0.45--0.72$\times$) because the search expands to cover most of the matrix, and the overhead of search traversal outweighs the savings from skipping cells. The optimal strategy depends on the divergence characteristics of the input sequences. In FILTR, users can explore different scheduling transformations to choose the best schedule for their data.

\paragraph{Pruning across different sequences.}
The choice of pruning heuristic is equally data-dependent. We evaluated three biological conditions that exhibit qualitatively different mutation profiles. For each condition, we computed the exact optimal alignment using full affine-gap scoring and compared it against three faster heuristics---banded, X-drop, and Z-drop---across a sweep of parameter settings, measuring both runtime and accuracy relative to the ground truth.

\begin{figure}
    \centering
    \includegraphics[width=1\linewidth]{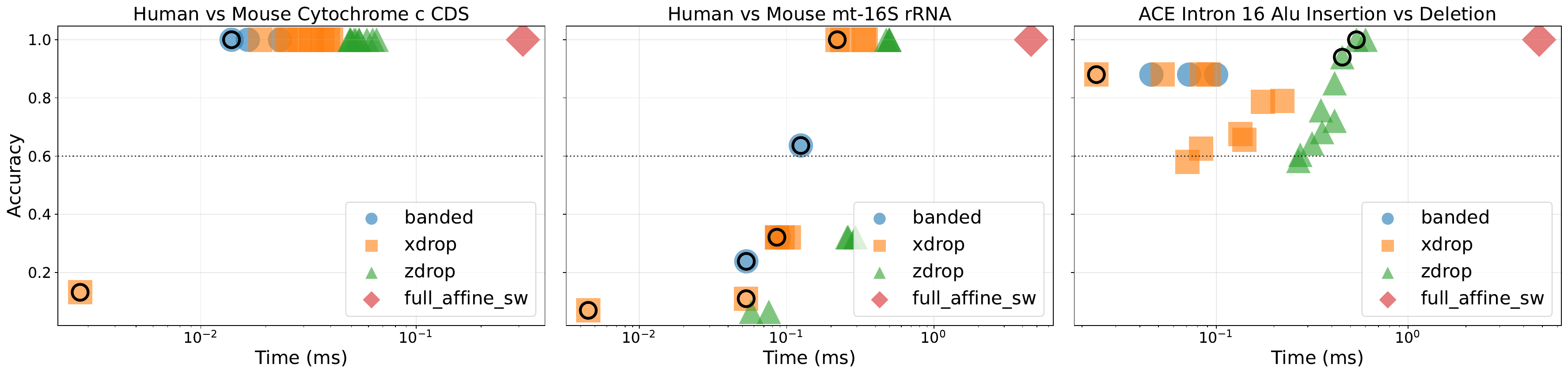}
  \caption{Accuracy vs. runtime for three biological conditions: resequencing (Human vs. Mouse Cytochrome c CDS, left), ortholog comparison (Human vs. Mouse mt-16S rRNA, center), and structural variation (ACE Intron 16 Alu Insertion vs. Deletion, right). The optimal heuristic differs by condition: banded is useful for resequencing, X-drop for ortholog, and Z-drop for structural variation. Each dot represents a different combination of bandwidth, X-drop, and Z-drop parameters; circles mark points on the Pareto frontier. Different applications have different accuracy requirements; the dotted line at 60\% marks a useful cutoff.}
  \label{fig:pareto-accuracy-time}
\end{figure}

\Cref{fig:pareto-accuracy-time} presents the resulting best option for each condition. In the resequencing case (left panel, Human vs. Mouse Cytochrome c CDS), narrow banded alignment achieves near-perfect accuracy at the lowest runtime, since mutations are small and the optimal path stays close to the main diagonal. For ortholog comparisons (center panel, Human vs. Mouse mt-16S rRNA), X-drop provides the best speed–accuracy tradeoff because its data-dependent band can widen to accommodate short bursts of insertion-deletions and then contract again when the alignment returns to a well-matching region. In the structural variation case (right panel, ACE Intron 16 Alu Insertion vs. Deletion), Z-drop is the only heuristic that maintains full accuracy even in the presence of long insertions, because it is more tolerant to extended gaps. With FILTR, users can easily plug in different pruning specifications and determine which pruning is best for their data.

\subsection{Transferring Heuristics}
\label{sec:transferring-heuristics}

\begin{figure}[t]
  \centering
  \includegraphics[width=0.9\textwidth]{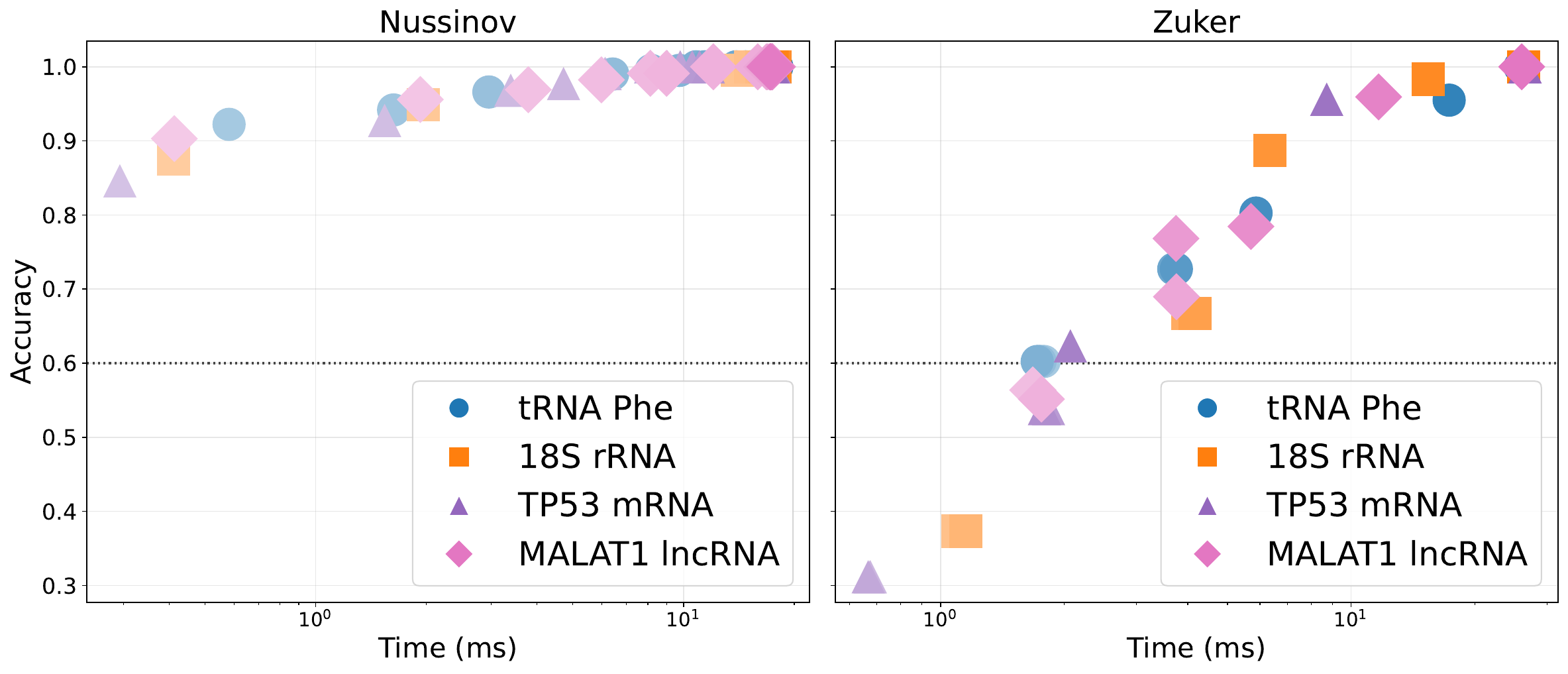}
  \caption{Each subplot shows a recurrence evaluated on four RNA sequences. Points represent different X-drop thresholds---Nussinov: 1–400 and unpruned; Zuker: 1–1000 and unpruned. Darker shades indicate larger X-drop values (less aggressive pruning). The dotted line marks an accuracy threshold of 0.6.}
  \label{fig:transfer-heuristics}
\end{figure}

FILTR enables users to construct new heuristics by recombining existing building blocks across problem domains. To illustrate this, we applied the X-drop pruning heuristic to three RNA secondary structure prediction recurrences. This experiment is not a rigorous evaluation of X-drop for RNA folding, but a demonstration of how FILTR's composable design makes it easy to transfer a pruning strategy to a new domain and assess whether it is worth exploring.

The Nussinov recurrence~\cite{nussinov1978} fills a triangular dynamic programming matrix indexed by subsequence endpoints $(i, j)$, iterated along increasing antidiagonals (\Cref{tab:dp_recurrences}). We reused the same four pruning constructs from alignment-based X-drop. We additionally modified the pruning program to include a warm-up period of $\text{antidiag}(i,j) \le 15$ (implemented with a conditional recurrence) during which pruning is disabled, since folding scores require several antidiagonals to become meaningful---unlike alignment, where even the first antidiagonal can produce nonzero scores. We extended this transfer to two additional RNA folding recurrences---Zuker~\cite{zuker1981} (minimum free energy folding) and McCaskill~\cite{mccaskill1990partition} (partition function folding).

The effectiveness of the transfer depends on the recurrence. The transfer to the McCaskill recurrence proved ineffective, likely because the recurrence aggregates contributions through multiplication and summation of exponentials rather than $\min$ or $\max$; a heuristic based on exponential contribution thresholds may be more suitable.

Nussinov and Zuker proved far more promising. \Cref{fig:transfer-heuristics} reports runtime and accuracy across four RNA classes (sequence lengths 76--500 nucleotides). For Nussinov, X-drop is remarkably effective: even at an aggressive threshold of $X=2$, accuracy remains perfect across all RNA classes, meaning pruning removes only cells that do not contribute to the optimal fold. This is likely because Nussinov scores grow monotonically and stay close to the optimum along the main antidiagonal, making it a natural fit for relative-score pruning. For Zuker, a tight threshold ($X=8$) prunes too aggressively (accuracy $0.3$--$0.6$), but a larger threshold ($X=600$) restores accuracy to $0.8$--$0.9$ while still providing modest speedups.

\section{Related Work}
\label{sec:Related-Work}
Hand-optimized sequence alignment libraries such as Ksw2 (used in minimap2~\cite{li2018minimap2}), Parasail~\cite{daily2016parasail}, and SeqAn~\cite{seqan} represent the dominant approach to achieving performance in biological sequence alignment. These systems implement dynamic programming recurrences directly in low-level code, embedding both the scoring semantics and the scheduling strategy within vectorized inner loops. Variants such as affine-gap~\cite{gotoh1982}, banded~\cite{ukkonen1985}, X-drop~\cite{altschul1997}, and Z-drop~\cite{li2018minimap2} are typically realized as separate pieces of code, each hand-engineered for a specific heuristic. Hardware acceleration efforts, including GPU-based implementations~\cite{liu2009cudasw} and FPGA-based efforts such as Darwin~\cite{darwin, koliogeorgi2021fpga} push this specialization further by synthesizing custom pipelines for alignment, achieving high throughput but sacrificing flexibility. The LOGAN framework~\cite{logan} illustrates this rigidity: each alignment heuristic requires explicit reconstruction of the compute pipeline. FILTR abstracts these implementations by representing scoring and pruning as schedulable constructs, allowing a single recurrence specification to generate optimized code for multiple heuristics and targets.


Compiler infrastructures for bioinformatics provide higher-level abstractions but have not generalized the structure of alignment recurrences themselves. Seq~\cite{seq2019} introduced a Python-like language for genomic pipelines that compiles to efficient code, but it targets data-flow orchestration rather than dynamic programming optimization. SARVAVID~\cite{sarvavid} adds distributed execution and workflow parallelism but still treats alignment algorithms as opaque calls. Researchers have also developed programming systems for recurrence problems. Dynamite~\cite{Birney1998Dynamite} compiles state-transition models into dynamic programming recurrences, but does not optimize its scheduling or pruning. The closest to our work are Recuma~\cite{recuma} and Bellman's GAP~\cite{bellmangap}. Recuma~\cite{recuma} formalized recurrence equations as a compilable intermediate representation, introducing a dependency–based scheduler that translates recurrence definitions to native code. However, Recuma assumes static domains and lacks primitives for pruning or dynamic domain restriction. Reptile~\cite{tariq2025reptile} is another recurrence compiler that extends Recuma with tiling optimizations for improved cache locality, but similarly does not support the class of optimizations needed for production bioinformatics algorithms. Bellman's GAP is a logic programming system whose focus is on reducing exponential search to polynomial recurrences. It does not support diagonal iteration or dynamic pruning, although it has limited support for pruning by placing a static band around the diagonal. FILTR extends this class of recurrence compilers by introducing domain-specific constructs, prune axes, score-based search, and shear domain transformations, which capture the irregular control flow of biological heuristics while preserving the analyzability and safety of recurrence transformations.

From the programming-languages perspective, FILTR builds on the separation of algorithm and schedule popularized by Halide~\cite{halide} and TACO~\cite{kjolstad2017taco}, and on loop optimization work done by the polyhedral community~\cite{feautrier1991,feautrier1992,lamport1974,banerjee1993}. Similar separation-of-concerns approaches appear in Lift~\cite{steuwer2017lift}, TVM~\cite{tvm2018}, and Tensor Comprehensions~\cite{vasilache2018tensorcomprehensions}. Halide demonstrated that schedules can independently control traversal, tiling, and fusion of dense image-processing loops, while TACO generalized this concept to sparse tensor formats through algebraic iteration graphs. FILTR adapts these principles to dynamic programming by defining the iteration space as a dependency-closed subset of the score and coordinate domain. Its iteration ordering and pruning directives serve a role analogous to Halide's scheduling primitives but operate on semantically constrained recurrence graphs, enabling the safe transformation of irregular, score-dependent dependencies.

\section{Conclusion}
\label{sec:Conclusion}
Hand-optimized bioinformatics algorithms achieve performance at the expense of hard-coding recurrence semantics and existing bioinformatics DSLs provide abstraction but not algorithmic flexibility. FILTR unifies these directions by bringing recurrence-level programmability and scheduling separation to biological recurrences, allowing heuristic variants to be expressed, composed, and compiled automatically.
Extending compilation to target GPUs and distributed execution is a natural next step. More broadly, FILTR's composable design creates a large search space of iteration orders, pruning strategies, and their combinations, but users must currently navigate this space manually. An autotuning layer that leverages the modularity of the three input languages to automatically discover the best configuration for a given dataset and hardware target is important future work.

\section{Data Availability}

The artifact for this paper includes the FILTR compiler. The frontend is embedded in Python, and the compiler itself is a Python repository that compiles inputs and generates C++ code. The artifact also includes scripts to reproduce the benchmarks in this paper, including the corresponding C++ kernels and the frontend code to generate them. The artifact will be submitted for artifact evaluation.

\newpage

\bibliographystyle{ACM-Reference-Format}
\bibliography{references}

% ...

\end{document}